\documentclass[aps,preprint,showpacs,superscriptaddress]{revtex4-1}  
\usepackage{graphicx}  
\usepackage{dcolumn}   
\usepackage{bm}        
\usepackage{amssymb}   
\usepackage{times}
\usepackage{amsmath}
\usepackage{amsfonts}
\usepackage{amsthm}
\usepackage[normalem]{ulem}

\usepackage{tabularx}
\usepackage{rotating}
\usepackage{multirow}
\usepackage{array}

\usepackage{fancyhdr}
\usepackage{graphicx}
\usepackage{epsfig}
\usepackage[dvipsnames]{xcolor}
\usepackage{tikz}
\usetikzlibrary{arrows,positioning,calc}
\usetikzlibrary{matrix}
\usetikzlibrary{shapes}
\usetikzlibrary{chains}

\usepackage{lipsum}
\usepackage[english]{babel}
\usepackage[utf8]{inputenc}
\usepackage[T1]{fontenc}
\usepackage{lmodern}
\usepackage{subfigure}
\usepackage{titlesec}
\usepackage{mathtools} 
\usepackage{tcolorbox} 
\usepackage{url}
\usepackage{epstopdf}

\newcommand{\ovl}[1]{\overline{#1}}
\begin{document}

\widetext


\title{Energy Storage in Steady States under Cyclic Local Energy Input}

\author{Y.Zhang}\email{yzhang@ichf.edu.pl}
\affiliation{Institute of Physical Chemistry, Polish Academy of Sciences, Kasprzaka 44/52, PL-01-224 Warsaw, Poland}
\author{R.Ho\l yst}\email{rholyst@ichf.edu.pl}
\affiliation{Institute of Physical Chemistry, Polish Academy of Sciences, Kasprzaka 44/52, PL-01-224 Warsaw, Poland}
\author{A.Macio\l ek}\email{ania.maciolek@gmail.com}
\affiliation{Institute of Physical Chemistry, Polish Academy of Sciences, Kasprzaka 44/52, PL-01-224 Warsaw, Poland}
\affiliation{Max-Planck-Institut f{\"u}r Intelligente Systeme, Heisenbergstr.~3, D-70569 Stuttgart, Germany}

\date{\today}

\begin{abstract}
We study periodic steady states of a lattice system under external cyclic energy supply using simulation. We consider different protocols for cyclic energy supply and examine the energy storage. Under the same energy flux, we found that the stored energy depends on the details of the supply, period and amplitude of the supply. 
Further, we introduce an adiabatic wall as internal constrain into the lattice and examine the stored energy with respect to different positions of the internal constrain. We found that the stored energy for constrained systems are larger than their unconstrained counterpart. 
We also observe that the system stores more energy through large and rare energy delivery, comparing to small and frequent delivery.  
\end{abstract}

\maketitle

\section{Introduction}
Equilibrium states of the many body systems are well established. They are described
by thermodynamics with its foundation justified by statistical mechanics. The central result
of statistical mechanics, \textit{i.e.}, the equilibrium probability distribution of a system in a heat
bath as given by the Boltzmann distribution, is considered to be the textbooks knowledge.
Such a general framework for non-equilibrium steady states does not exist yet.

Nonetheless, different approaches that address specific  non-equilibrium situations have been developed.  
For non-equilibrium systems that fall under linear response regime, \textit{i.e.}, equilibrium systems under small perturbations,  phenomenological  arguments led to the formulation of the fluctuation-dissipation theorem (FDT)~\cite{toda02, kubo, marconi}.  
These systems  can also be described by linear irreversible thermodynamics, which  provides  an expression for the entropy production in the form of affinities and fluxes \cite{groot1962, pottier}. 
Further developments concern the extension of  FDT to  non-equilibrium steady states (NESS).  Various forms of FDTs were proposed, which link the response to a small perturbation from the NESS to the  correlation functions in the NESS \cite{marconi, udo-fdt, baiesi}. 
For systems in which fluctuations are eminent, results now known as the fluctuation theorems (FT) were proved, which are valid also for deep NESSs. 
These theorems establish restrictions on the probability distribution of quantities defined on  trajectories \cite{revUdo, van2015ensemble}.
Different FTs have been proved for a variety of systems such as chaotic  \cite{gallavotti1995dynamical} or  stochastic systems \cite{lebowitz1999gallavotti}, and for various processes such as diffusion \cite{kurchan1998fluctuation} or relaxation  \cite{evans1994equilibrium}. 
The FTs for the probability distribution of work,  were proved by Jarzynski and Crook \cite{jarzynski,crooks2000path}. 
By generalizing the concept of thermodynamic quantities on the trajectory level, these relations can be unified under the framework of stochastic thermodynamics \cite{revUdo}. 
They are particularly relevant for small systems such as molecular motors or nano-devicies \cite{bustamante2005nonequilibrium,schmitt2015molecular}, chemical reaction systems \cite{rao2016nonequilibrium} and systems with strong fluctuations \cite{ciliberto2017experiments}. 
The central results of contemporary non-equilibrium thermodynamics are critically reviewed in Ref.~\cite{Marsland_2017}.

Here, we address the problem of  energy storage in NESSs. It is interesting  and relevant for potential applications, e.g., in energy harvesting~\cite{PhysRevLett.102.080601},   to know how the average 
energy storage depends on details of the energy supply and  on internal geometrical constrains of a system. The similar question was  raised recently  in Ref~\cite{holyst2019flux}.
The approach used in that paper is based on  two quantities: the energy $\Delta U = U -U_0$ stored in non-equilibrium states,  and the total energy flux  $J_U$ in these states. $U_0$ is  the internal energy at the equilibrium state, obtained after the shutdown of the energy input. In Ref.~\cite{holyst2019flux}, $\Delta U$ was determined for an ideal gas and for the  Lennard-Jones fluid.
It was found that  the stored energy depends not only on the total energy flux, $J_U$, but also on the protocol of the energy transfer into the system.
For three different protocols of energy transfer, it was demonstrated that $\Delta U/J_U = \mathcal{T}$ is minimized in the stationary states formed in these systems.
Minimization is with  respect to all constrained states of the system, similar as in equilibrium thermodynamics.
$\mathcal{T}$ has an interpretation of the characteristic time scale of energy outflow from the system immediately after the shutdown of the energy input.
It was demonstrated that  $\mathcal{T}$ is minimized in stable states of the Rayleigh-Benard cell.

Following this methodology,  
we examine the stored energy in an Ising lattice system under different protocols of local periodic energy input using numerical simulation.
By introducing internal constrains, we  test the hypothesis of the minimization of $\Delta U/J_U$. 
We perform simulation in two spatial dimensions using the so-called 'deterministic  Ising algorithm' introduced by Creutz~\cite{creutz, creutz-02, creutz-03},  
which allows  to perform local energy input and to measure local temperature.   Contrary to the conventional Monte Carlo simulation with Metropolis algorithm, this
algorithm permits study of non-equilibrium phenomena, such as heat flow. On the other hand, it has
been shown to behave in almost the same way as Monte Carlo simulations of the original Ising model~\cite{aktekin2000simulation}. 

This approach is different from what is often used for  non-equilibrium lattice systems.
The paradigm lattice model for studying non-equilibrium phenomena is the driven lattice gas model~ \cite{zia, zia-02}. It   can be treated both  analytically and by standard Monte Carlo simulations because the concept of stochastic Markov processes with discrete states comes natural for lattice systems. 
Examples of driven lattice gas models are the asymmetric simple exclusion process (ASEP), totally asymmetric exclusion process (TASEP) and their variations. With relatively simple rules, these models can already demonstrate complex non-equilibrium phenomena, such as, spontaneous symmetry breaking, non-equilibirum phase separation and current fluctuations \cite{mukamel2000phase, mukamel-01, mukamel-02, gingrich}. 
External drivings can be achieved, for example, by attaching two reservoirs which induce transport of energy or particles through the system or by
assigning a biased transition rate in the preferable direction \cite{zia}. 
In contrast  to the present case, in these models  the temperature  is often assumed to be homogeneous, and the external drives are uniformly applied throughout the system.
 
The Ising lattice system under the local energy input was studied in Ref.~\cite{Li_2012}. However, in that paper the focus was on the non-equilibrium phase diagram 
and the energy input was realized by dividing the system into two sectors  in contact with thermal baths
with markedly different temperatures. 
In our model,  following the 'deterministic Ising algorithm',  each spin is assigned with a dynamic variable (originally referred  to as the demon). The spin can exchange energy with this demon, which serves as a local heat bath and the temperature can be extracted from its mean energy. This method allows us to input energy specifically to each spin by elevating the energy of its demon and measure the resulting temperature distribution.  Although the Creutz’s dynamics is deterministic, the statistical nature of the whole system arises from
the complexity of a large phase space. This is similar to the molecular dynamics approach, which solves
deterministic evolution of a system described by Hamiltonian, i.e., it does not use random numbers. Previously, this algorithm has been used, among others, to simulate the electrocaloric effect in alloys~\cite{ponomareva2012bridging}, to study  dynamics of discrete systems with long range interactions \cite{lederhendler2010long}, or  thermal conductivity in lattice  systems  without \cite{saito,harris,neek2006monte,muglia2012dynamical,casartelli2007heat} and with an interface \cite{PhysRevE.97.052141,PhysRevE.99.052128}, where the lattice is typically attached to two heat baths at different temperatures. 
Our lattice system is in a homogeneous heat bath. We input energy internally and allow the heat to flow through the boundary. Since energy can only dissipate through the boundary, the temperature inside the system is not homogeneous. In experiments, such systems can be realized by shining a laser onto a material or a solution. 

The paper is organized as follows. In section 2, we describe the model and the simulation method. We define different protocols of energy input and measure the corresponding stored energy. In section 3, we introduce an internal constrain into the system. We measure the same quantities now with respect to different positions of the constrain. We conclude in section 4.

\section{Ising model with local energy input}
\subsection{Model description}
\begin{figure}
\centering
\includegraphics[scale=0.48]{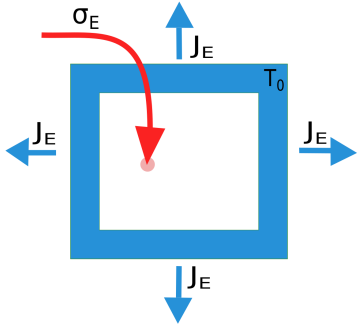}
\caption{Scheme of an unconstrained system. A two dimensional open lattice system, denoted in white in the center, is surrounded by a heat bath at temperature $T_{0}$ denoted in blue (grey). External energy $\sigma_{E}$ is supplied into the system. Energy currents are denoted as $J_{E}$.}
\label{scheme_sys}
\end{figure}

We consider a 2-dimensional open  lattice system of size $L \times L$ in a heat bath of temperature $T_{0}$. Each point of the lattice  is occupied by a spin of value $\pm 1$. The scheme of this system is shown in Fig.\ref{scheme_sys}.
 
The system is driven to non-equilibrium states with externally controlled energy supply $\sigma_{E} (\vec{r},t)$,  which depends both on position $\vec{r} = (x,y)$ and time $t$. 
The external energy supply is achieved through the direct elevation of local energy, which results in local temperature elevation. 
The energy is dissipated through the boundary of the system and the current is denoted by $\vec{J}_{E}(\vec{r})$. 
From energy conservation, it follows that at steady states the amount of energy supplied into the system equals the amount of energy dissipated out. Therefore, the total energy flux $J_{U}$ is given by
\begin{equation}
J_{U} \equiv  \oint_{S} \vec{J}_{E}(\vec{r}) d\vec{S} = \int_{V} \sigma_{E}(\vec{r}) d \vec{r}, 
\label{tot_flux}
\end{equation}
where $\vec{S}$ is the boundary of the system with direction that is pointing outward. 
Hence, the total energy flux can be measured from the total energy supply per time (over the whole system). 
We would like to point out that even under homogeneous energy supply, that is, if $\sigma_{E}(\vec{r},t) = \sigma_{E}$, due to a finite rate of energy dissipation, the steady state temperature profile $T(\vec{r})$ is not homogeneous.

We define different protocols of energy supply through $\sigma_{E}(\vec{r}, t)$. 
We focus on a periodic supply with period $\tau$.
 The system reaches periodic steady states where quantities averaged through a period remain constants 
\cite{rikvold1998kinetic,tome1990dynamic}. Examples of periodic steady states driven by external magnetic fields can be found in \cite{chakrabarti1999dynamic}. In these states, instead of quantity $O(t)$, its corresponding average over one or several periods $\overline{O}$ are evaluated. 
For instance, we define the averaged energy flux as
\begin{equation}
\overline{\sigma_{E}}(\vec{r}, \tau) \equiv \dfrac{1}{\tau} \int_{t}^{t+\tau} \sigma_{E}(\vec{r}, t) dt,  
\label{aver_sigma}
\end{equation}
and the averaged system energy as 
\begin{equation}
\ovl{U}_{\mathrm{sys}}(\tau) \equiv \dfrac{1}{\tau} \int_{t}^{t+\tau} U(t) dt,   
\end{equation}
with the (averaged) stored energy 
\begin{equation}
\Delta U(\tau) = \ovl{U}_{\mathrm{sys}}(\tau) - U_{0}, 
\end{equation}
where $U_{0}$ is the system energy in equilibrium.

\subsection{Simulation method}
As stated previously, classical Monte Carlo simulation is not suitable for simulations involving local energy supply.  Monte Carlo dynamics usually assumes a homogeneous temperature profile. This is no longer the case for our system. In this section, we will briefly introduce the 'deterministic Ising algorithm' developed by Creutz~\cite{creutz, creutz-02, creutz-03}. We then describe our method. In practice, our system is a combination of the Metropolis algorithm for exchanging energy with the heat bath, and the deterministic Ising algorithm for the energy supply and internal energy diffusion. 

\subsubsection{Update procedure}
In the algorithm proposed by Creutz, an extra variable $D(\vec{r})$ is associated with each spin $s(\vec{r})$ at $\vec{r}$. 
The local Hamiltonian at time $t$ is given by
\begin{equation}
\mathcal{H} (t) = -Js(\vec{r},t)\sum_{\vec{r}'}s(\vec{r}',t) + D(\vec{r},t), 
\end{equation}
where $J$ is the interaction strength between spins, $\vec{r}'$ represents the neighbors of the lattice point $\vec{r}$. For $\vec{r} = (x,y)$, $\vec{r}' \in \{(x,y\pm 1), (x\pm 1, y)\}$. 
$D(\vec{r})$ can exchange energy with $s(\vec{r})$ and  $D(\vec{r})/J = 0,1,2,\cdots$.  This variable plays the role of the kinetic energy of the local spin $s(\vec{r})$ in that the temperature can be defined through measuring the average of the kinetic energy, in analogy  to the Molecular Dynamics simulation. $D(\vec{r})$ can also be regarded as a heat bath with only one degree of freedom. Assuming that the energy exchange is fast between $s(\vec{r})$ and $D(\vec{r})$, the spin is in equilibrium with $D(\vec{r})$ and have the same temperature $T(\vec{r})$. 
The energy distribution of $D(\vec{r})$ is expected to follow the Boltzmann distribution, $P(D(\vec{r})) \propto \exp(-D(\vec{r})/k_{\textrm{B}}T(\vec{r}))$. Throughout the paper,   temperature is measured in units of $J/k_{\mathrm{B}}$ and  the Boltzmann constant $k_{\textrm{B}}$ is set to $1$.
 Local temperature can be determined through measuring the expectation value of $D(\vec{r})$ \cite{creutz, creutz-02} 
\begin{equation}
\langle D(\vec{r},t) \rangle \equiv \dfrac{\sum_{0}^{\infty} n\exp(-n/T(\vec{r},t))}{\sum_{0}^{\infty} \exp(-n/T(\vec{r},t))} = \dfrac{1}{\exp(1/T(\vec{r},t)) - 1} , 
\label{ave_D}
\end{equation}
where $\langle \cdot \rangle$ is the ensemble average and $n=0,1,2,\cdots$. Rewriting eqn.(\ref{ave_D}), we have  
\begin{equation}
T(\vec{r},t) = \dfrac{1}{\ln (1/\langle D(\vec{r},t) \rangle + 1)}. 
\label{T(r)}
\end{equation}

\begin{figure}
\centering
\includegraphics[scale=0.48]{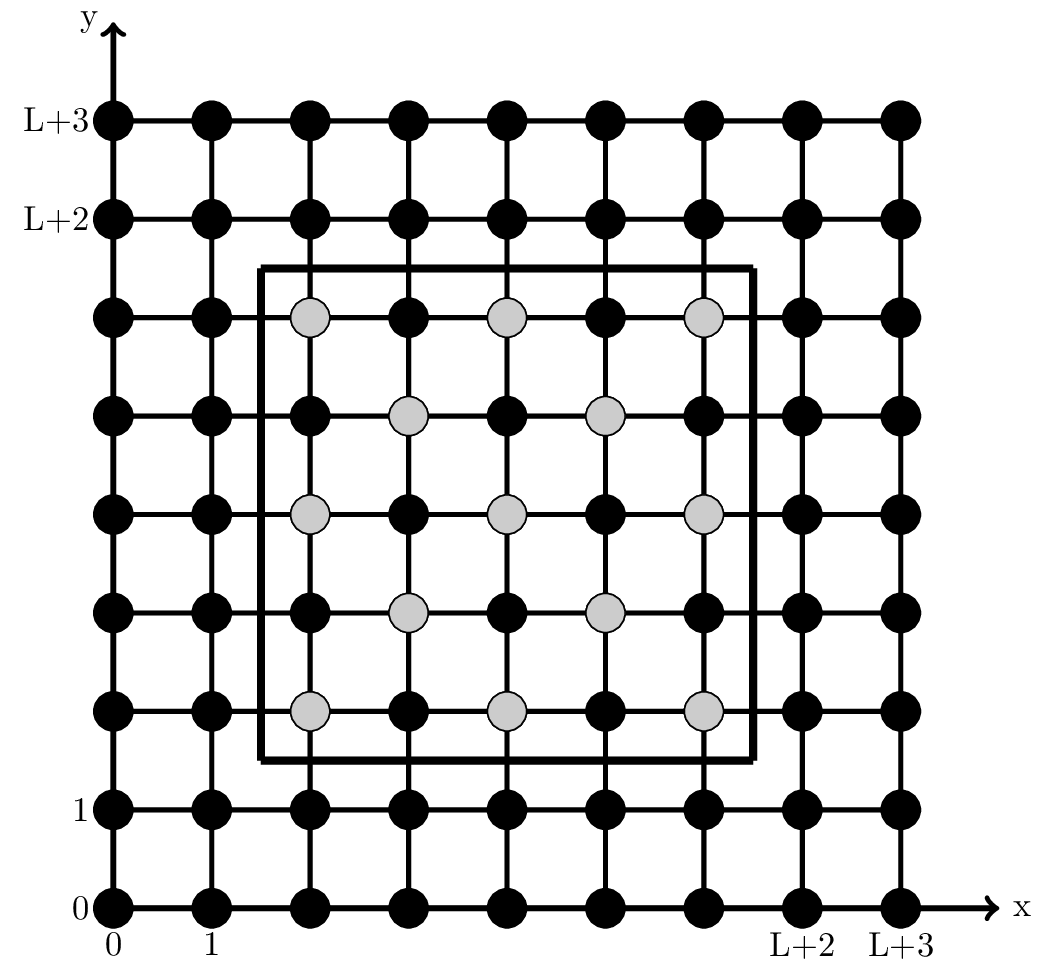}
\caption{Scheme of the check-board update of the lattice system. System of interest are inside the thick black lines and  is surrounded by two layers of spins representing the heat bath. Black and grey dots of the system denote two group of spins during the check-board update. For a system of size $L \times L$, $x, y$ ranges from $0$ to $L+3$. }\label{creutz-system}
\end{figure}

An update in the deterministic dynamics is performed under the rule that the local energy is preserved. 
Details of each update are as follow. When a spin is chosen, we first calculate local energy difference under spin flip using 
\begin{equation}
\Delta E = 2Js(\vec{r})\sum_{\vec{r}'} s(\vec{r}').  
\label{dE}
\end{equation}
If the flip lowers the local energy $\Delta E \leq 0$, or if the momentum counterpart can provide enough energy for this update, that is, if $D(\vec{r},t) - \Delta E \geq 0$, the flip is accepted and $D(\vec{r},t+1) = D(\vec{r},t) - \Delta E$. Otherwise, this flip is rejected. 

In our simulation, we combine the Metropolis dynamics \cite{newman1999} with the deterministic dynamics. Spins in the system are separated into two parts, two layers of spins that serves as the heat bath (heat bath spins) surround spins that serves as the system (system spins). For example, for an open system of size $L \times L$, we use $(L+4) \times (L+4)$ number of spins in the simulation.  
Heat bath spins are updated using the Metropolis dynamics with transition rates at temperature $T_{0}$. System spins are updated using the deterministic dynamics explained above. This set-up is similar to the set-up in \cite{saito}, except that our system is an open system. 

The full procedure is as follows. 
In each Monte Carlo step, we first update all the heat bath spins using the Metropolis dynamics. These spins are selected sequentially. After a spin $s(\vec{r})$ is selected, a random number $\gamma \in (0,1)$ is generated and the local energy difference under the spin flip is calculated using eqn.(\ref{dE}). If $\Delta E \leq 0$ or if $\gamma \leq \exp(-(\Delta E/T_0))$, then the spin is flipped $s(\vec{r}) = -s(\vec{r})$. Otherwise, $s(\vec{r})$ remains the same. 

Then we update system spins using the deterministic dynamics described previously. In order to ensure that all spins are updated at the same time, that is, to avoid updating a spin using its updated neighbors, the spins are not chosen sequentially, rather in a check-board manner. A scheme of this check-board is shown in Fig.~(\ref{creutz-system}). The Monte Carlo step from configuration ${s(\vec{r}, t)}$ to ${s(\vec{r}, t+1)}$ is separated into two half-steps. During the first half-step, either grey spins or black spins are updated. The rest are updated during the second half-step.

\subsubsection{Energy supply realization}
\begin{figure}
\centering
\subfigure[]{\includegraphics[scale=0.48]{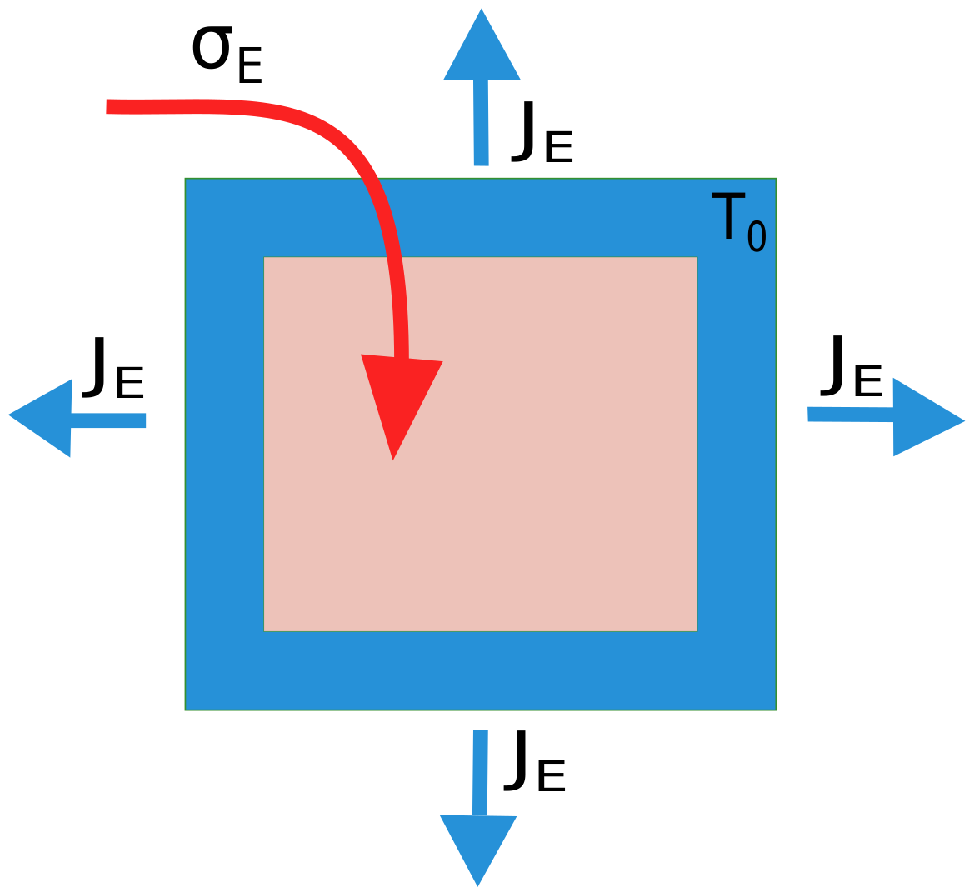}} 
\hspace{-.7cm}
\subfigure[]{\includegraphics[scale=0.48]{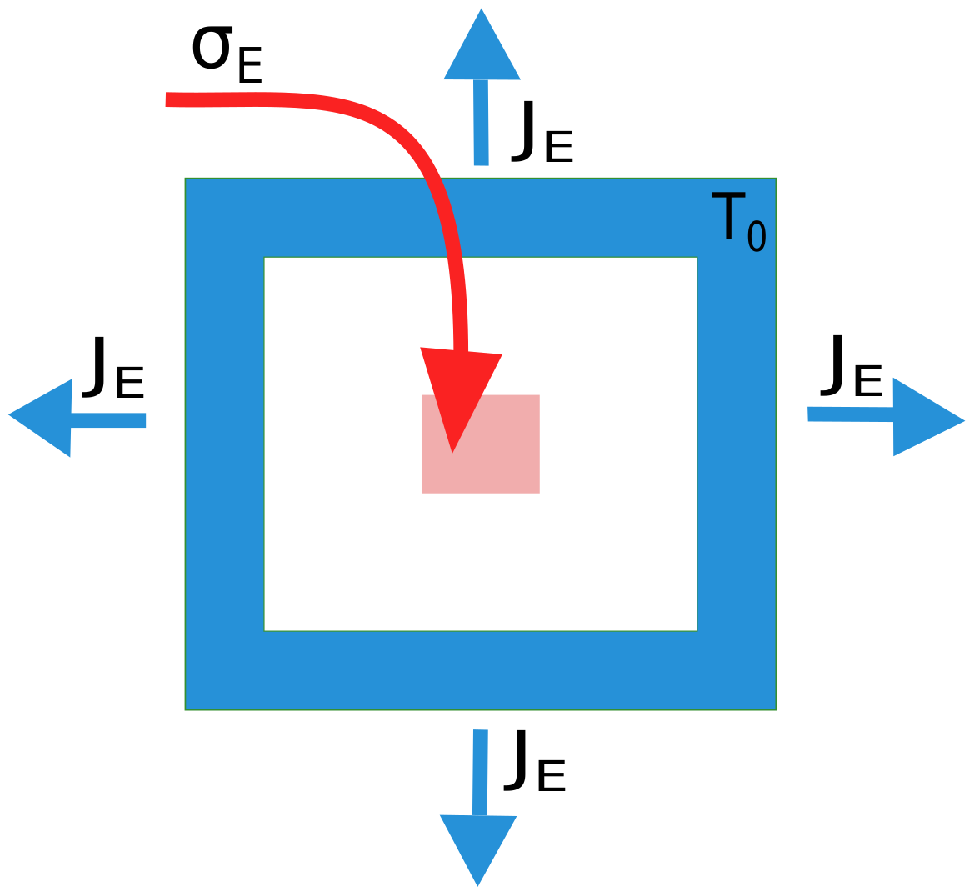}} 
\hspace{-.7cm}
\subfigure[]{\includegraphics[scale=0.48]{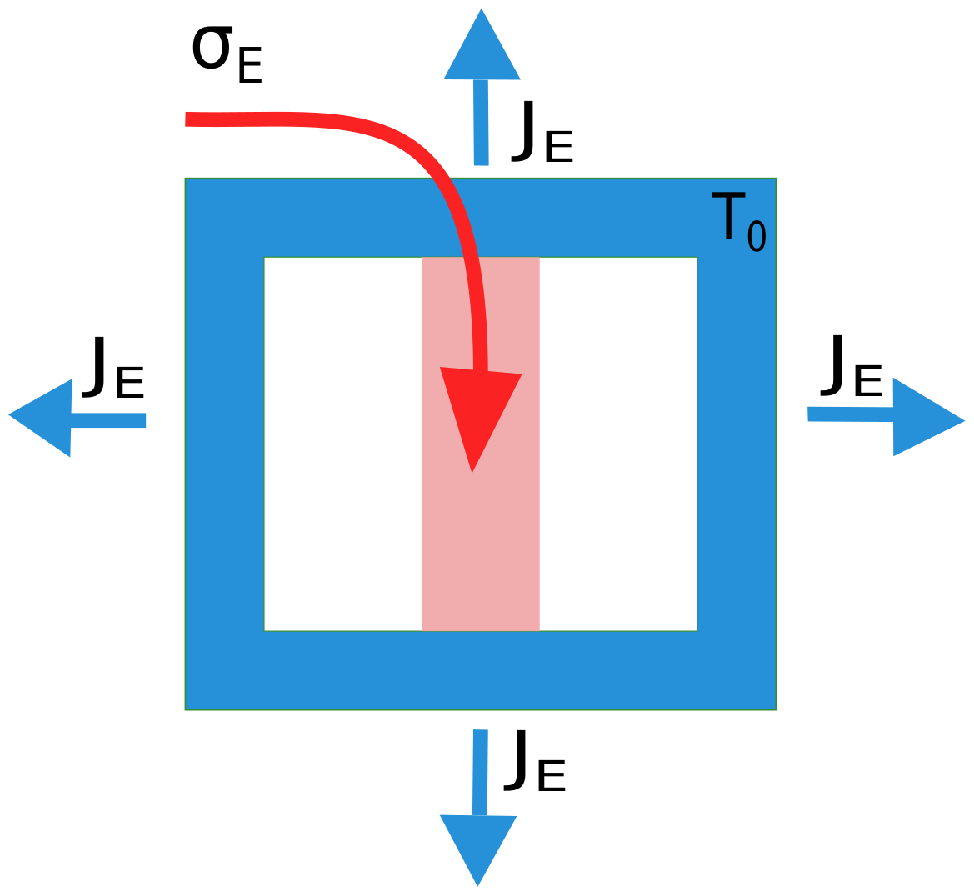}}
\caption{Schemes of the three geometries $A_{1,2,3}$ used for the  energy supply. Area of each geometry is shaded in red (light grey). $(a) A_{1}$ spans the whole system. $(b) A_{2}$ is square of size $10 \times 10$ at the center of the system. $(c) A_{3}$ is a stripe of size $L \times 4$ in the middle of the system.  }
\label{scheme_sigmaE}
\end{figure}

Local energy supply is realized by directly elevating the variable $D(\vec{r},t)$ with $\sigma_{E}(\vec{r},t)$. 
The explicit form of $\sigma_{E}(\vec{r},t)$ can be written as
\begin{equation}
\sigma_{E}(\vec{r},t) = E_{0}h(t) X_{A_{i}}(\vec{r}), 
\end{equation}
where $E_{0}$ is the amplitude (in units of $J$) and $h(t)$ is a periodic function, $h(t) = h(t+\tau)$. The simplest case of a periodic supply is to supply every $\tau$ time units (Monte Carlo steps). This can be represented using the delta function, $h(t) = \delta(t,n\tau)$, where $n=1,2,3,\cdots$. $X_{A_{i}}(\vec{r})$ is an indicator function that has value $1$ at the lattice sites where energy is supplied. In this paper, we use three types of geometry $A_{1,2,3}$ (shown in Fig.\ref{scheme_sigmaE}). $A_{1}$ covers the whole system, hence is a homogeneous input. $A_{2}$ is a square of size $10 \times 10$ at the center of the system. $A_{3}$ is a stripe of size $L \times 4$ in the middle of the system. These geometries can be represented as follow 
\begin{equation}
\begin{aligned}
\mathrm{case\ 1} \quad A_{1} &= \{\vec{r}|\forall \vec{r}\}, N_{1} = L^{2}\\
\mathrm{case\ 2} \quad A_{2} &= \{\vec{r}|L/2-4 \leq x \leq L/2+5, L/2-4 \leq y \leq L/2+5\}, N_{2}=100\\
\mathrm{case\ 3} \quad A_{3} &= \{\vec{r}|L/2-1 \leq y \leq L/2+2\}, N_{3}=4L. 
\end{aligned}
\end{equation}
where $N_{i}$ is the number of spins (or area) in geometry $A_{i}$, $i = 1,2,3$. Using definitions from eqn.(\ref{tot_flux}) and eqn.(\ref{aver_sigma}), we represent the energy input protocol of geometry $A_{i}$ as $J_{U}(A_{i}, E_{0}, \tau)$ and the averaged flux is 
\begin{equation}
\ovl{J_{U}}(N_{i}, E_{0}, \tau) = \dfrac{E_{0}}{\tau}N_{i}. 
\end{equation}
The averaged flux through the system depends on three variables, energy amplitude, the period and the area of the geometry,. 

\subsubsection{Measurement}
\begin{figure}
\centering
\includegraphics[scale=0.3]{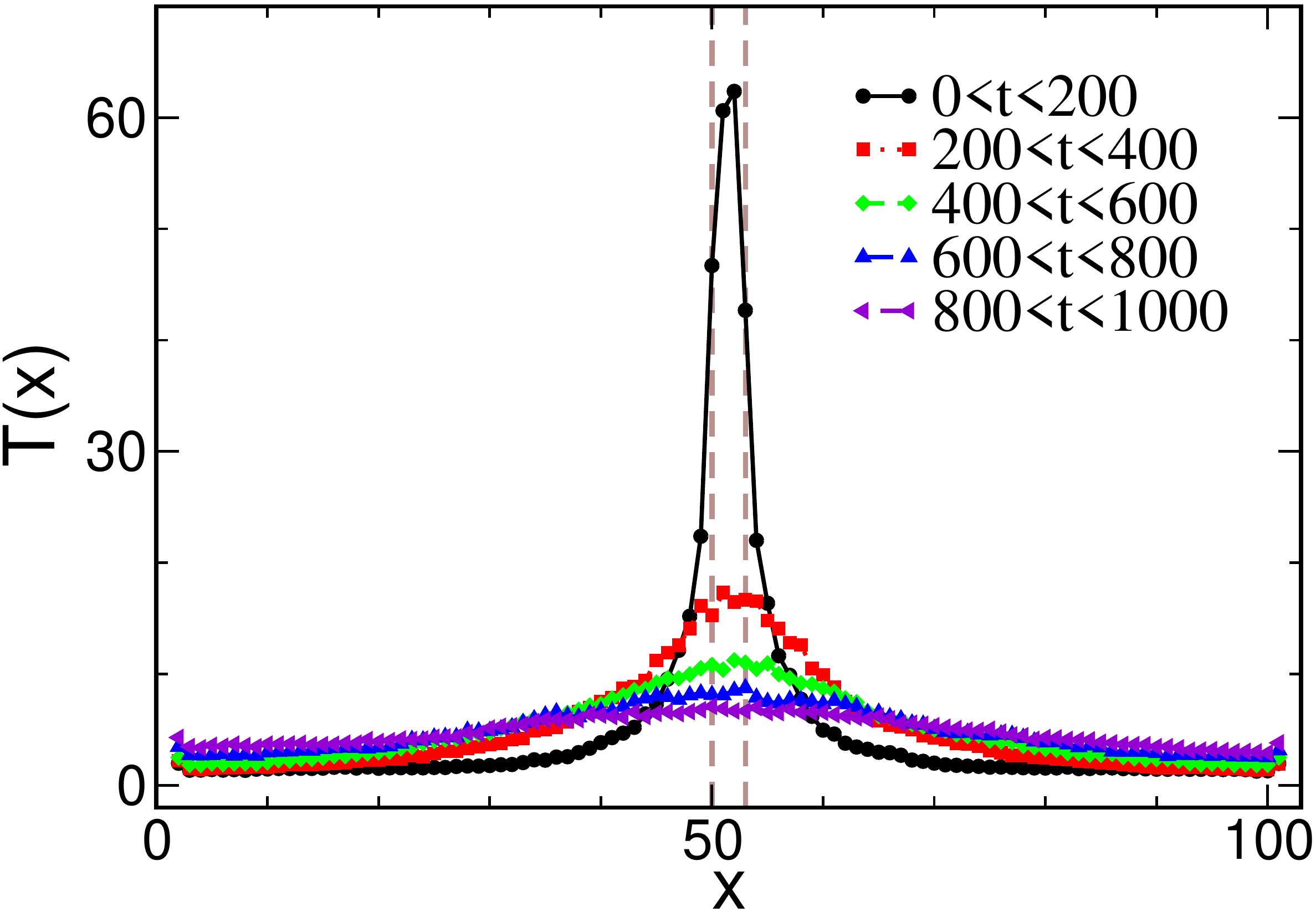}
\caption{Averaged column temperature profiles under energy supply protocol  $J_{U}(A_{3}, 100, 1000)$. The whole period  is divided into time intervals of 200 MC steps over which  the temperature profiles  are  averaged over these time intervals.
The $x$-coordinate of the stripe subjected to the energy supply lies between two vertical dashed lines. Temperature is measured in units of  $J/k_{\mathrm{B}}$. 
}
\label{Txt}
\end{figure}

The total energy of the system is
\begin{equation}
U_{\mathrm{sys}}(t) = -J\sum_{\langle \vec{r}, \vec{r}' \rangle}s(\vec{r},t)s(\vec{r}',t) + \sum_{\vec{r}}D(\vec{r},t), 
\end{equation}
where $\langle \vec{r}, \vec{r}' \rangle $ are neighboring lattice points. We set the interaction strength $J$ equal  to $1$. In steady states, the temperature profile is measured using eqn.(\ref{T(r)}). In simulation, 
the ensemble average is obtained through average over realizations
\begin{equation}
\langle D(\vec{r},t) \rangle = \dfrac{\sum_{M=1}^{N} D_{M}(\vec{r},t)}{N}, 
\end{equation}
where $D_{M}(\vec{r},t)$ is the local demon energy at time $t$ from realization $M$ and $N$ denotes the total number of realizations. Typically in our simulation $N$ ranges from $100$ to $500$.
For periodic steady states, we define averaged temperature profile $\ovl{T}(\vec{r}, \tau)$ through  
\begin{equation}
\ovl{T}(\vec{r}, \tau) = \dfrac{1}{\ln \big(1/\ovl{D}(\vec{r}, \tau) +1 \big) }, 
\end{equation}
where $\ovl{D}(\vec{r}, \tau)$ is obtained from $\int_{t}^{t+\tau}\langle D(\vec{r},t) \rangle /\tau$. 
Within each period the temperature profile changes with time. An example of how the temperature profile relaxes within a period in such periodic steady states is shown in Fig.\ref{Txt}. Under adiabatic conditions (no energy flux outside the system) and using geometry $A_{3}$, we supply energy at time $0$ with $E_{0}=100$. We average $T(\vec{r},t)$ over each column and over every $200$ steps. 
As we can see, this average relaxes and tends to equilibrate within the system. 

\subsection{Simulation results}
\begin{figure}
\centering
\subfigure[]{\includegraphics[scale=0.5]{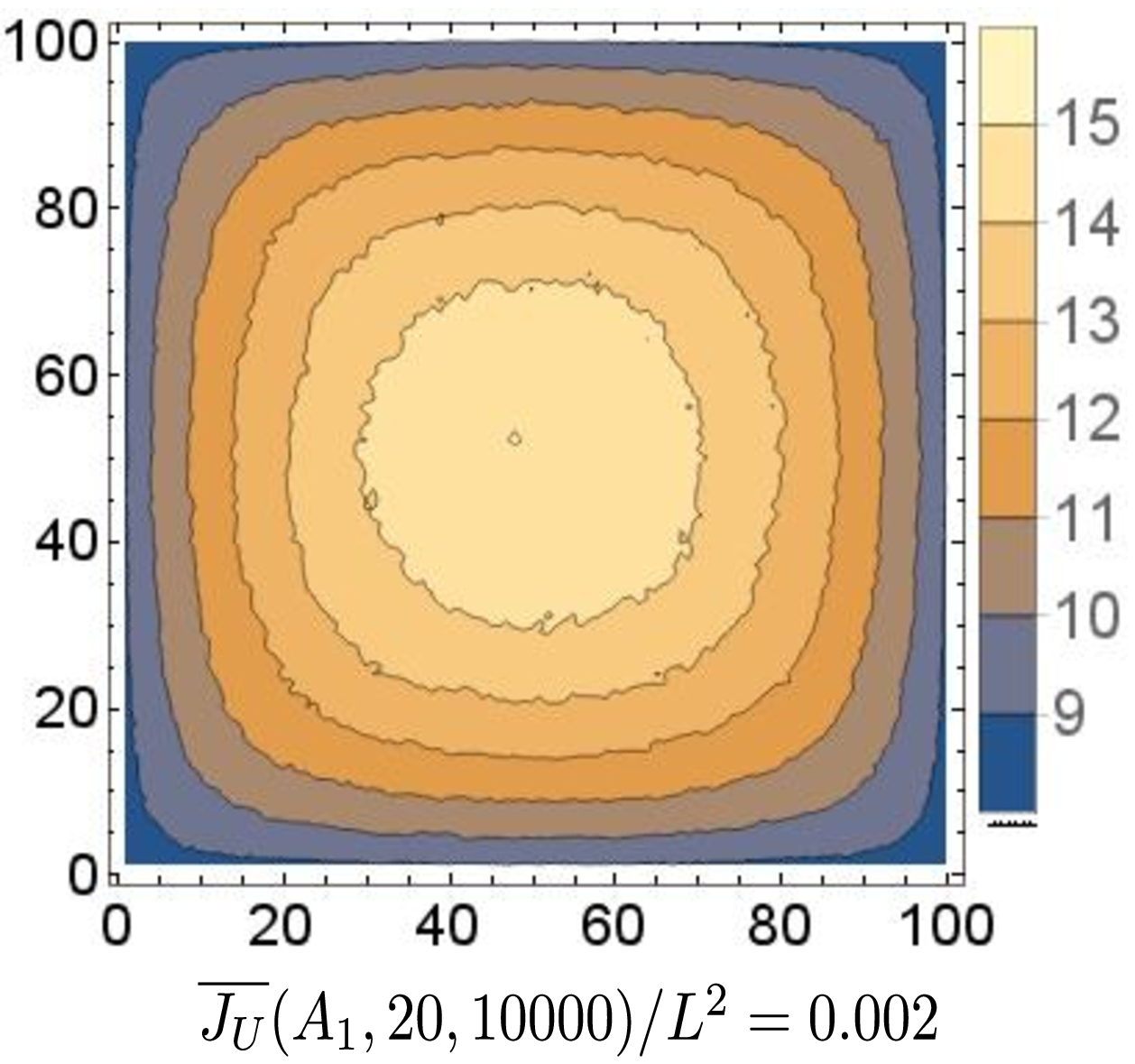}}
\subfigure[]{\includegraphics[scale=0.5]{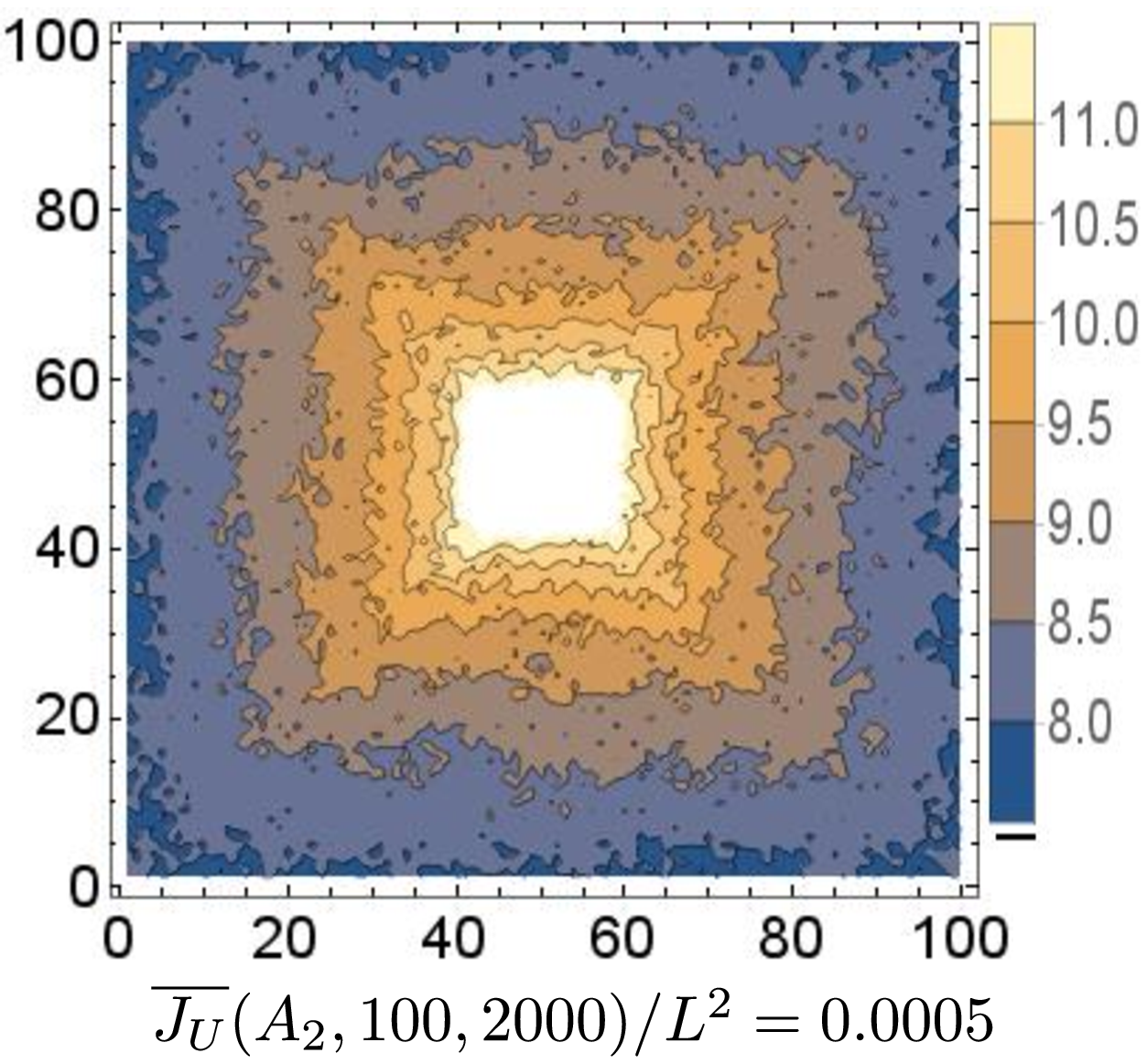}}
\subfigure[]{\includegraphics[scale=0.5]{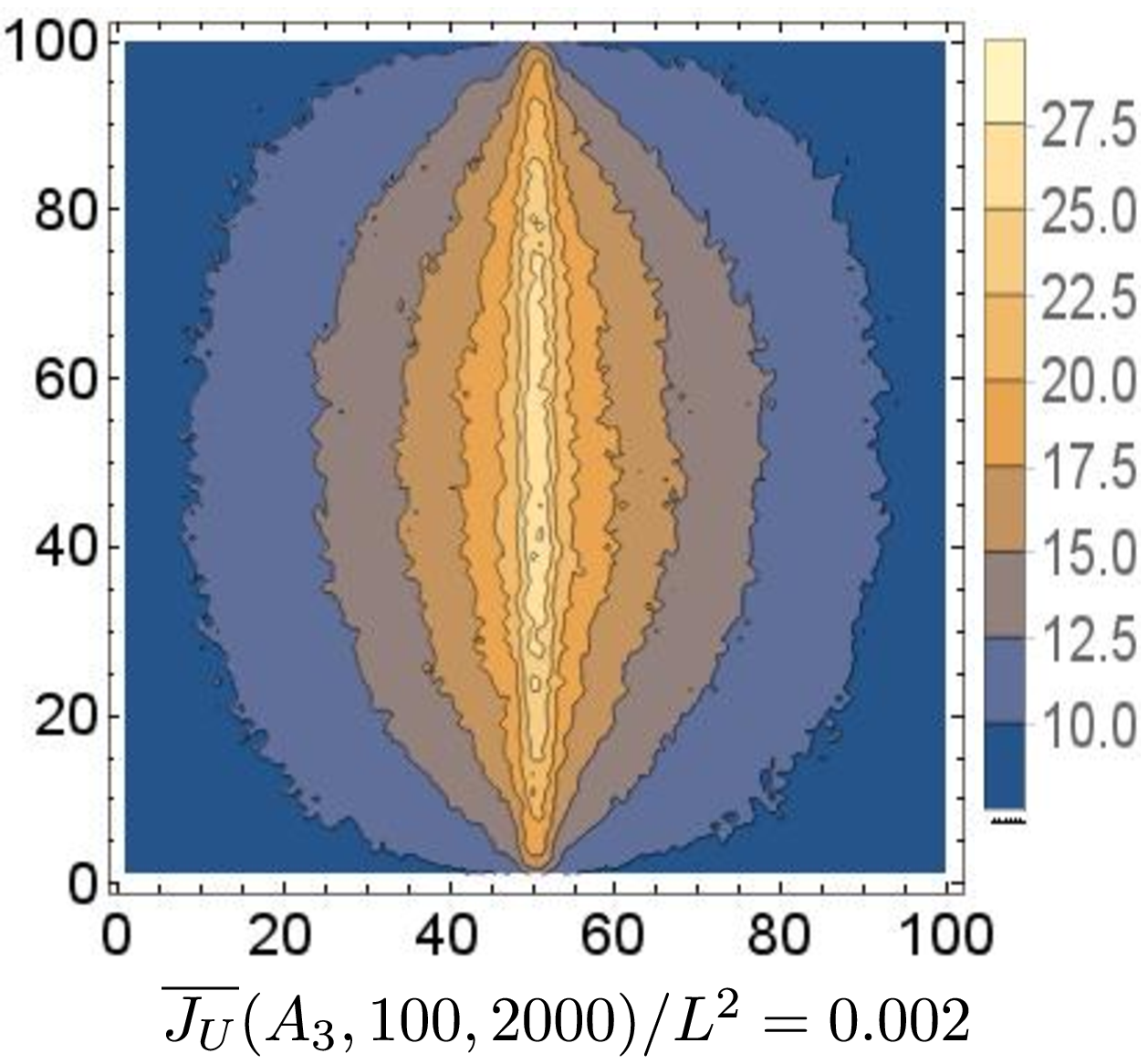}}
\caption{Contour plots of averaged temperature profiles $\ovl{T}(\vec{r}, \tau)$ under 
different geometries shown in Fig.\ref{scheme_sigmaE}. Corresponding protocols and averaged fluxes for energy supply are denoted under each graph. Temperature is measured in units of  $J/k_{\mathrm{B}}$. The white area in the centre of $(b)$ denotes temperature higher than the shown scale. }
\label{Txy_unconstrain}
\end{figure}

\begin{figure}
\centering
\subfigure[]{\includegraphics[scale=0.3]{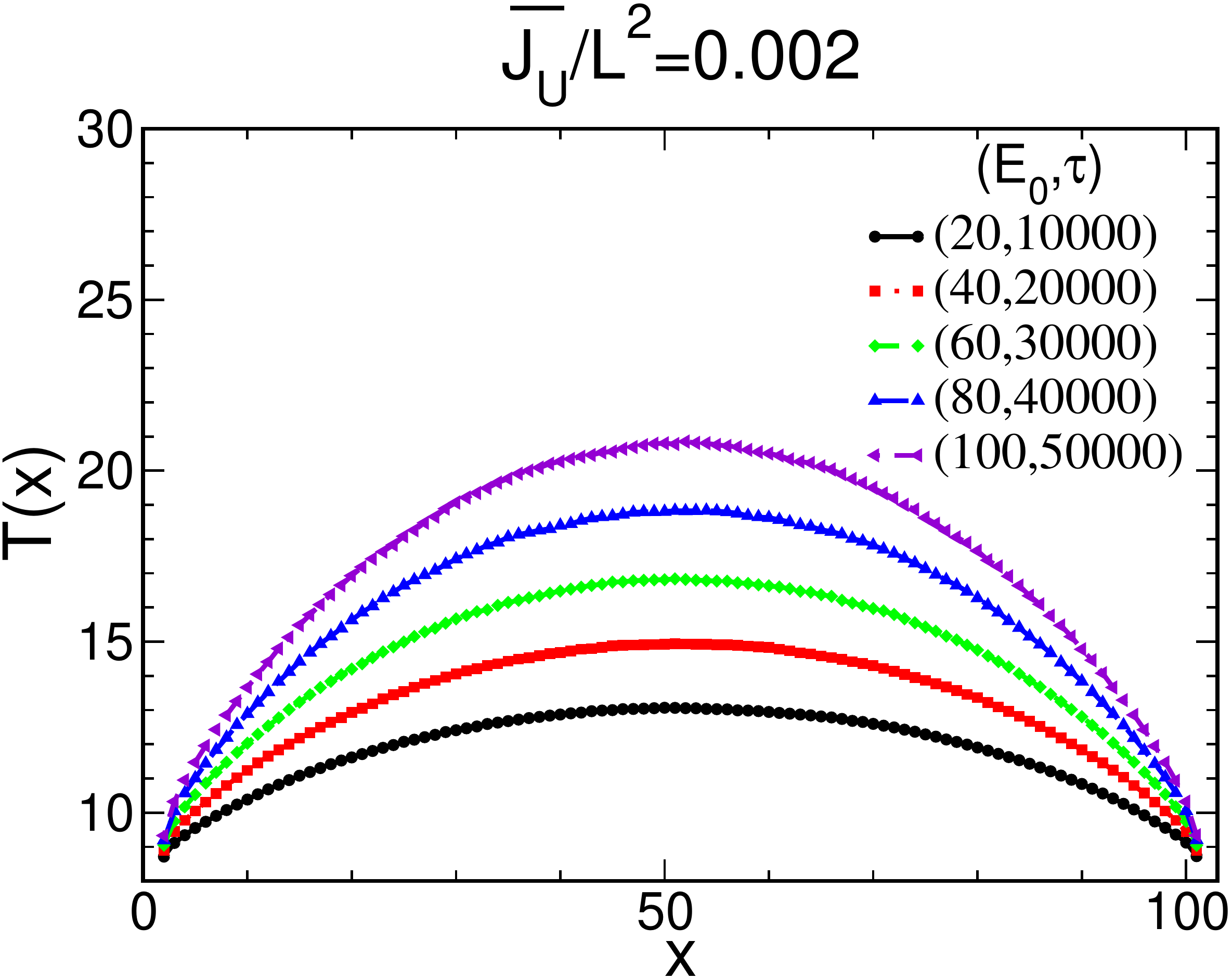}\label{Tx_1}}
\subfigure[]{\includegraphics[scale=0.3]{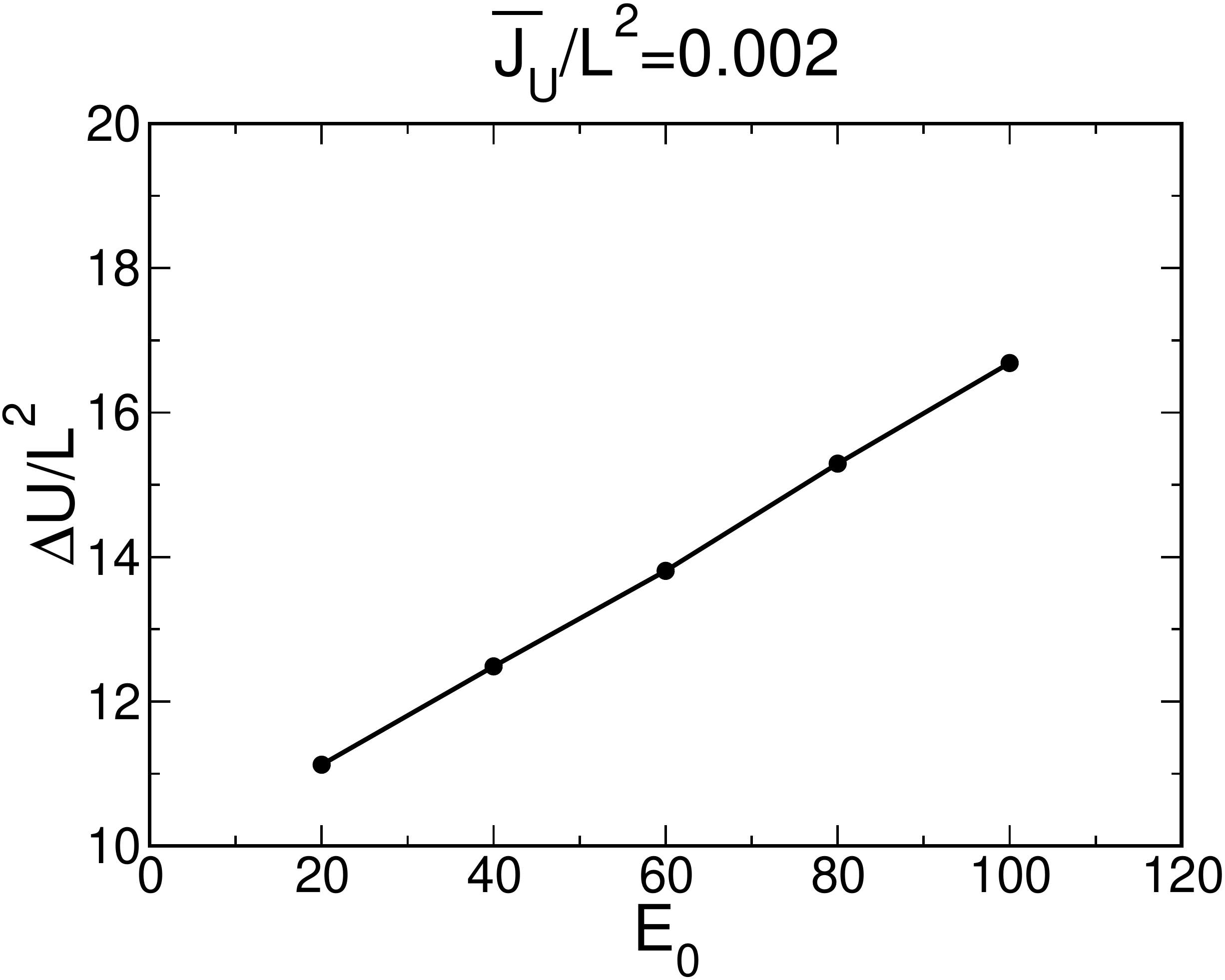}\label{dU_1}}
\subfigure[]{\includegraphics[scale=0.3]{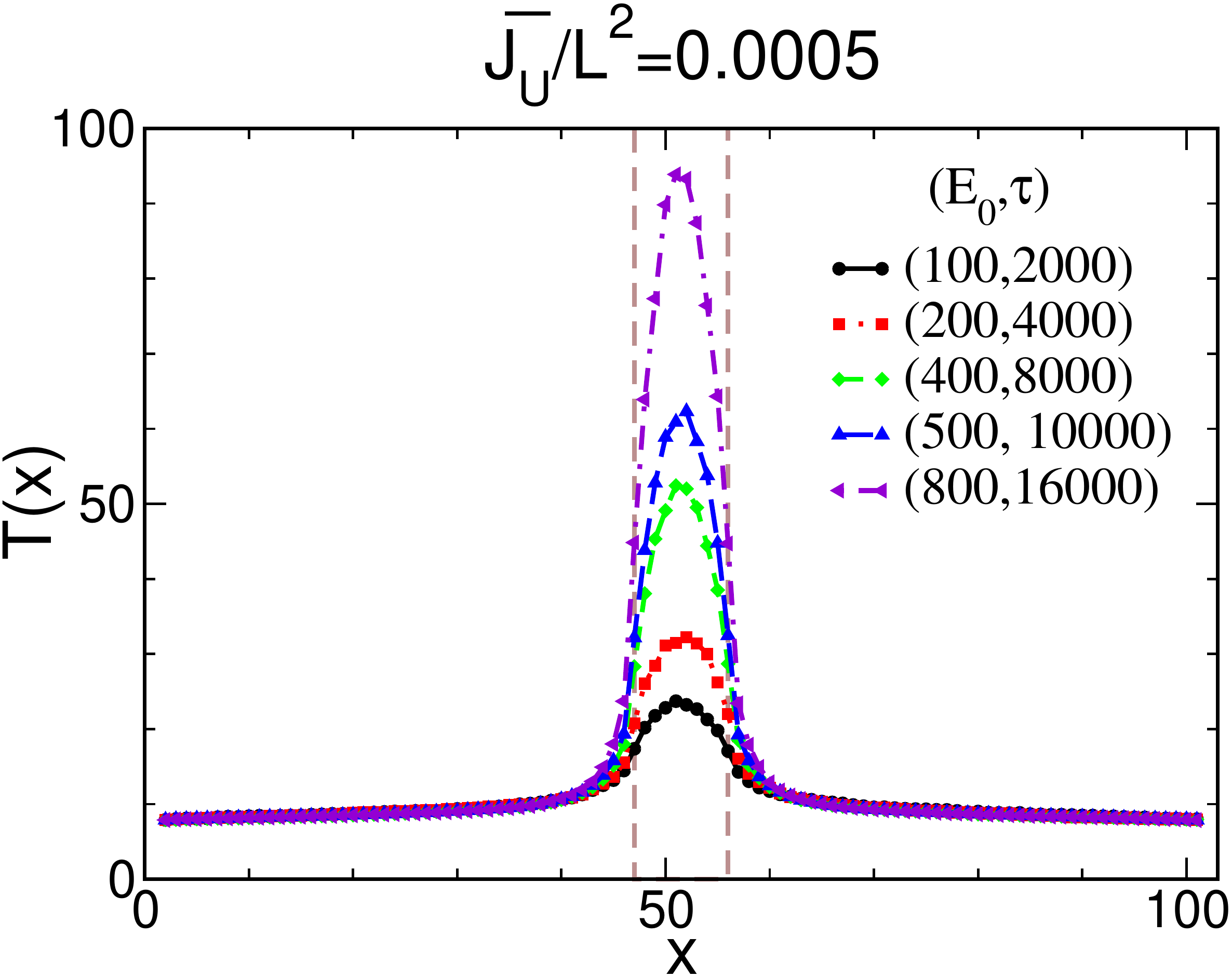}\label{Tx_2}}
\subfigure[]{\includegraphics[scale=0.3]{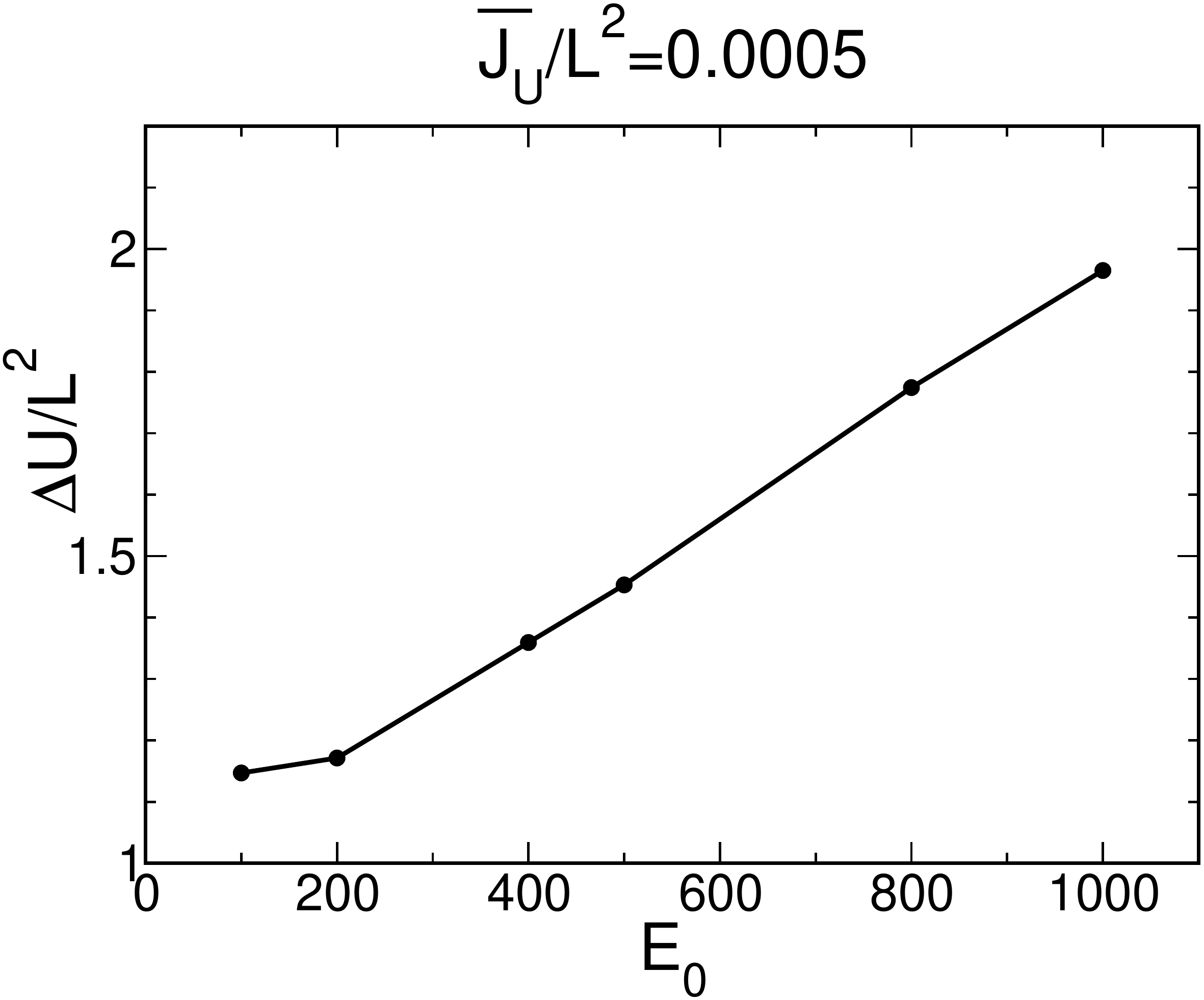}\label{dU_2}}
\subfigure[]{\includegraphics[scale=0.3]{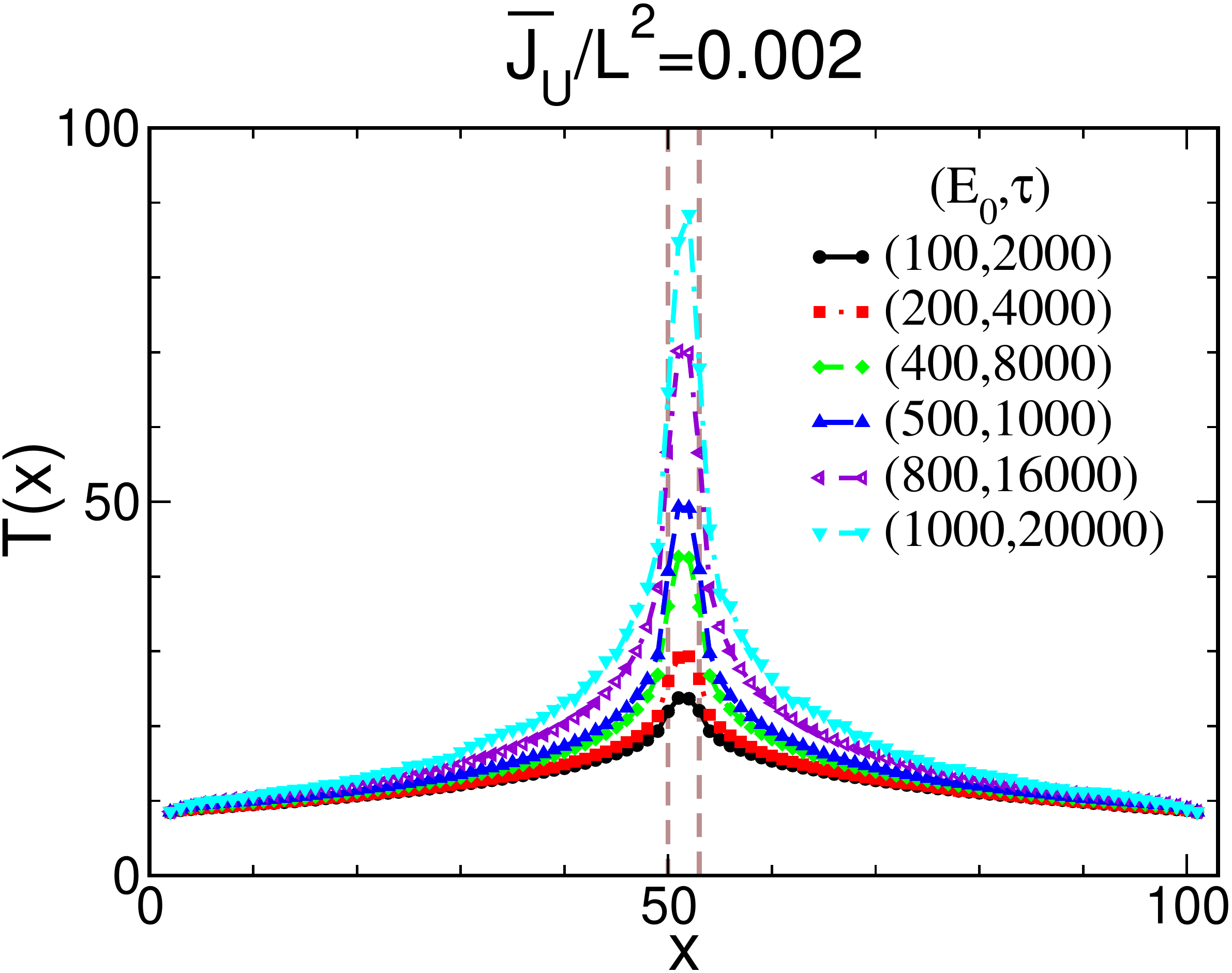}\label{Tx_3}}
\subfigure[]{\includegraphics[scale=0.3]{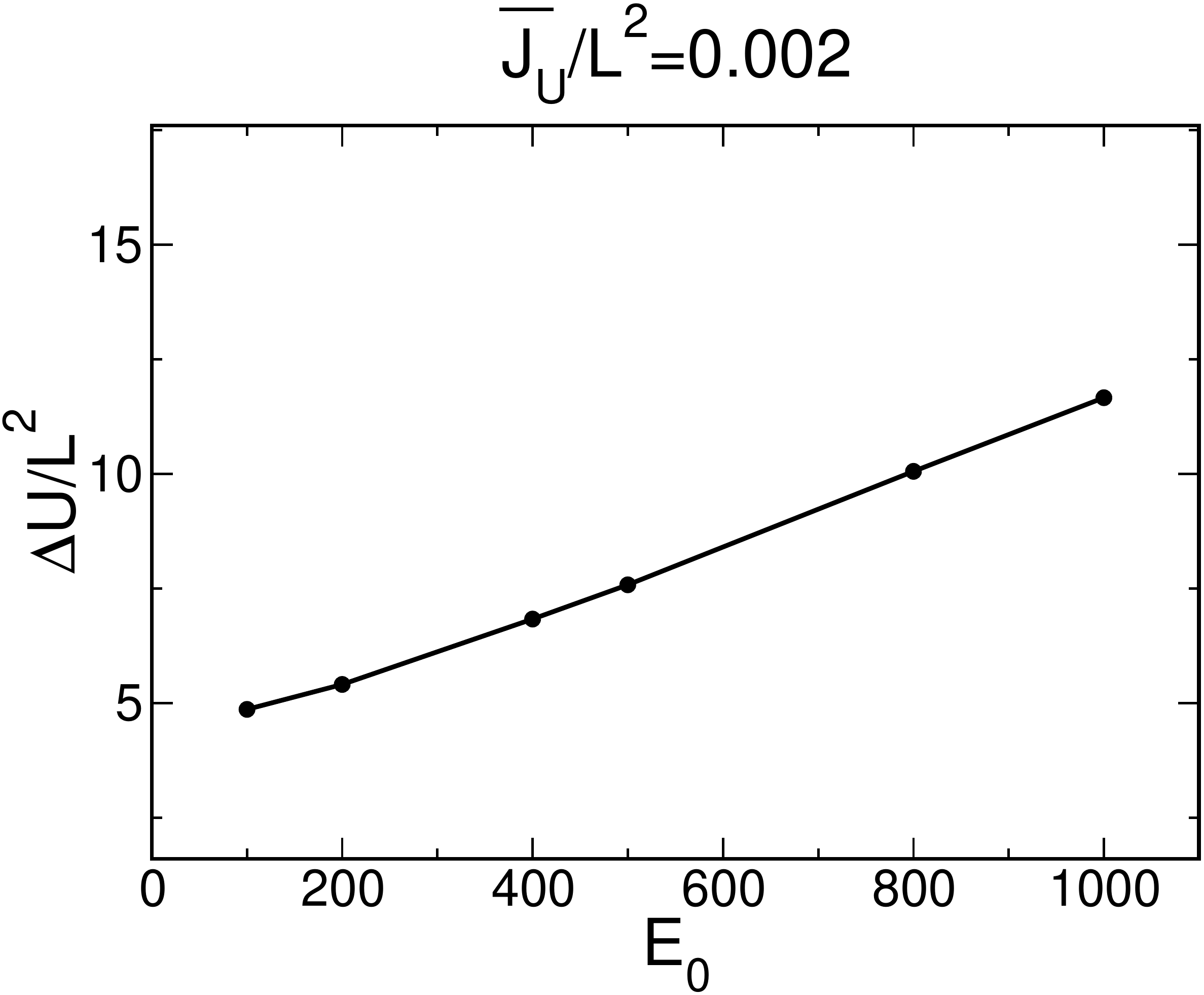}\label{dU_3}}
\caption{$(a)(c)(e)$ Column temperature profiles $\ovl{T}(x, \tau)$ under averaged constant energy flux $\ovl{J_{U}}$ for different geometries. The geometry and the $\ovl{J_{U}}$ per spin are denoted under each panel where specific values of the $(E_{0}, \tau)$ pair are denoted for each curve. Temperature is measured in units of  $J/k_{\mathrm{B}}$. 
$(b)(d)(f)$ Energy storage $\Delta U$ per spin under constant $\ovl{J_{U}}$ for different geometries. 
}
\label{Tx_unconstrain}
\end{figure}

First, we compute the stored energy under different protocols. 
Upon fixing the geometry $A_{i}$ and the amount of total flux $\ovl{J_{U}}$, we compare the stored energy $\Delta U$ under different pairs of $(E_{0}, \tau)$. 
Examples of contour plots of the averaged temperature profile $\ovl{T}(\vec{r}, \tau)$ under different geometries are shown in Fig.\ref{Txy_unconstrain}. For easier demonstration, we further define the averaged column temperature as $\ovl{T}(x, \tau) \equiv \sum_{y}\ovl{T}(x,y, \tau) /L$. 
Results of $\ovl{T}(x, \tau)$ are shown in Fig.\ref{Tx_1}, \ref{Tx_2} and \ref{Tx_3}. Each panel shows results of the column temperature from a single geometry. Different curves in each panel correspond to various $E_{0}$ and $\tau$ such that $\ovl{J_{U}}$ is fixed.  The stored energy $\Delta U$ per spin for each geometry are shown in Fig.\ref{dU_1}, \ref{dU_2} and \ref{dU_3}. 

As we can see from the figure, different protocols of energy supply lead to different steady states, demonstrated by different temperature profiles and different stored energy.
Further, in each geometry, we find that the energy stored from large but rare deliveries (large $E_{0}$ and $\tau$) is greater than that from small but frequent deliveries (small $E_{0}$ and $\tau$).
This holds if the characteristic time scale $\mathcal{T}=\Delta U/J_U$  of energy outflow from the system is smaller than the  period $\tau$ of the energy supply.
 For $\mathcal{T}$  comparable to $\tau$ the stored energy is minimal.  For $\mathcal{T}$ larger than $\tau$, the system does not reach periodic steady states.

\section{Constrained system}
Now, we consider the conjecture  formulated in Ref.~\cite{holyst2019flux}, which states that the
quantity $\Delta U/J_{U}$ is minimized at steady states. This conjecture has been verified in several systems. Following similar methodology, we introduce an adiabatic horizontal wall as the internal constrain into the lattice system. By adjusting the positions of the constrain, we compare temperature profiles and stored energy under different supply protocols.  

\begin{figure}
\centering
\includegraphics[scale=0.5]{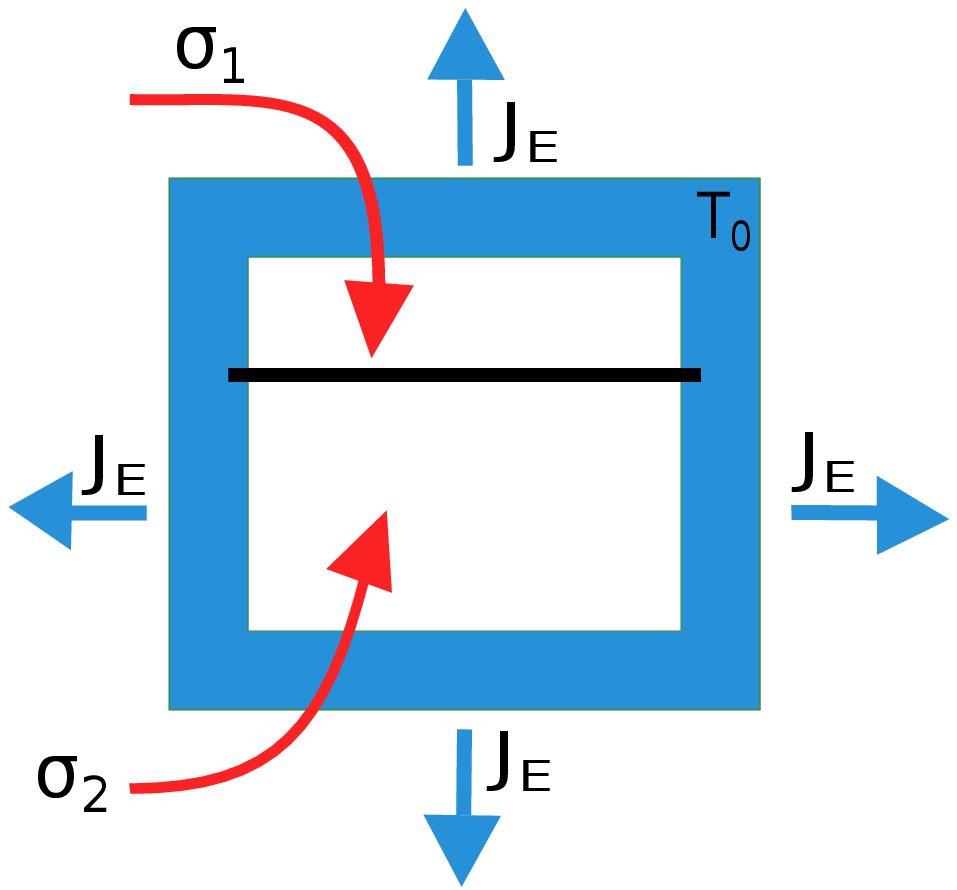}
\caption{Scheme of a constrained system. The horizontal line denotes the constrain which is an adiabatic wall. Energy flux is non-zero in each subsystem. }
\label{scheme_sys_con}
\end{figure}

An adiabatic horizontal wall is placed between row $y_{w}$ and $y_{w}+1$. It separates the system into two subsystems: upper and lower. A scheme of this constrained system is shown in Fig.\ref{scheme_sys_con}. 
In practice, this is achieved by setting the interaction strength between spins at row $y_{w}$ and $y_{w}+1$ to zero. This constrain does not change the total amount of energy influx, $\textit{i.e.}$, $\ovl{J_{U}} = \ovl{J_{U}}_{\mathrm{upper}} + \ovl{J_{U}}_{\mathrm{lower}}$. We use geometry $A_{1}$ and $A_{3}$ (Fig.\ref{scheme_sigmaE}). This insures that when adjusting $y_{w} \in (2,L+1)$ each subsystem will have a nonzero heat flux. 

\begin{figure}
\centering
\subfigure[]{\includegraphics[scale=0.25]{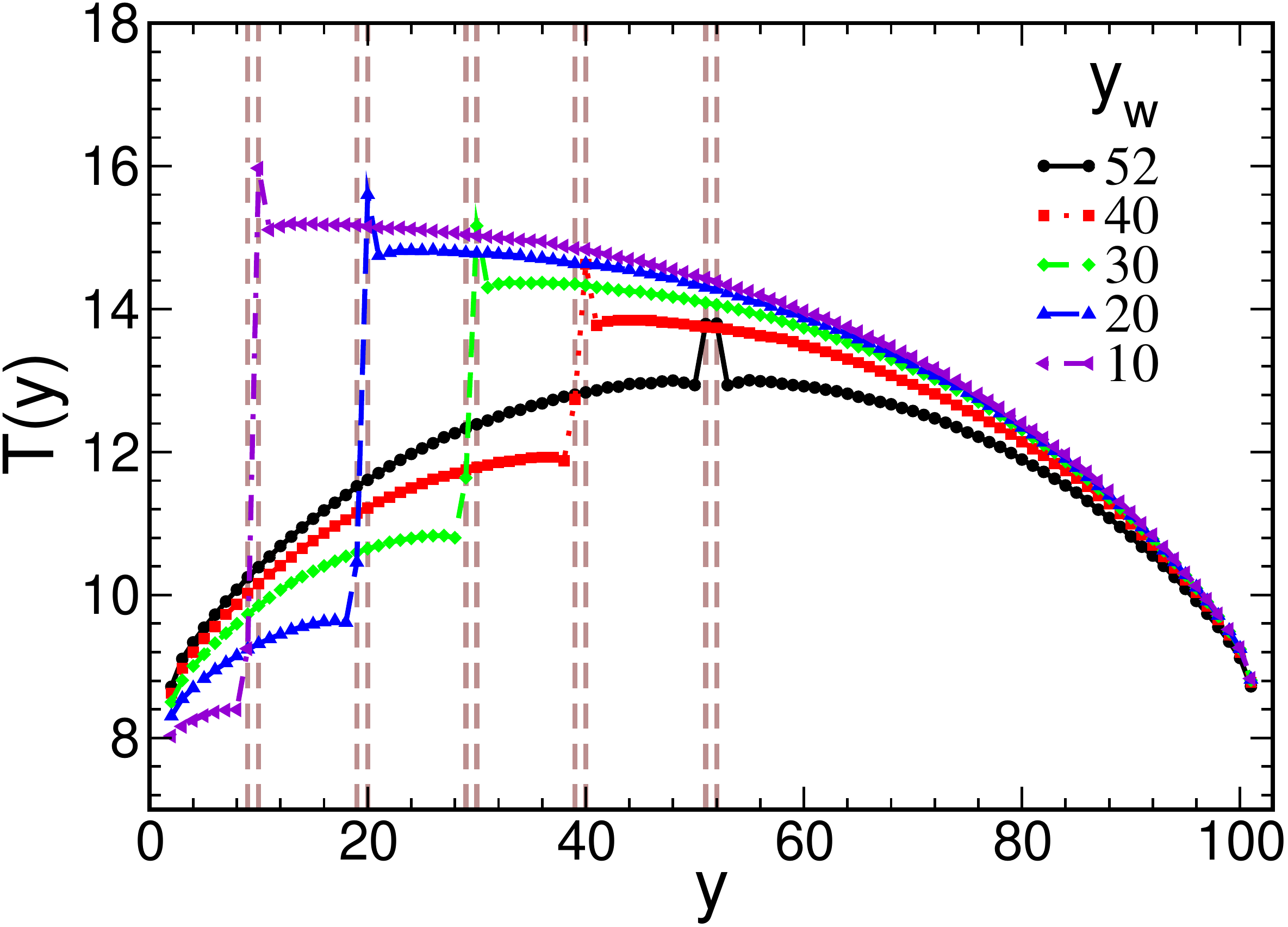}}
\subfigure[]{\includegraphics[scale=0.25]{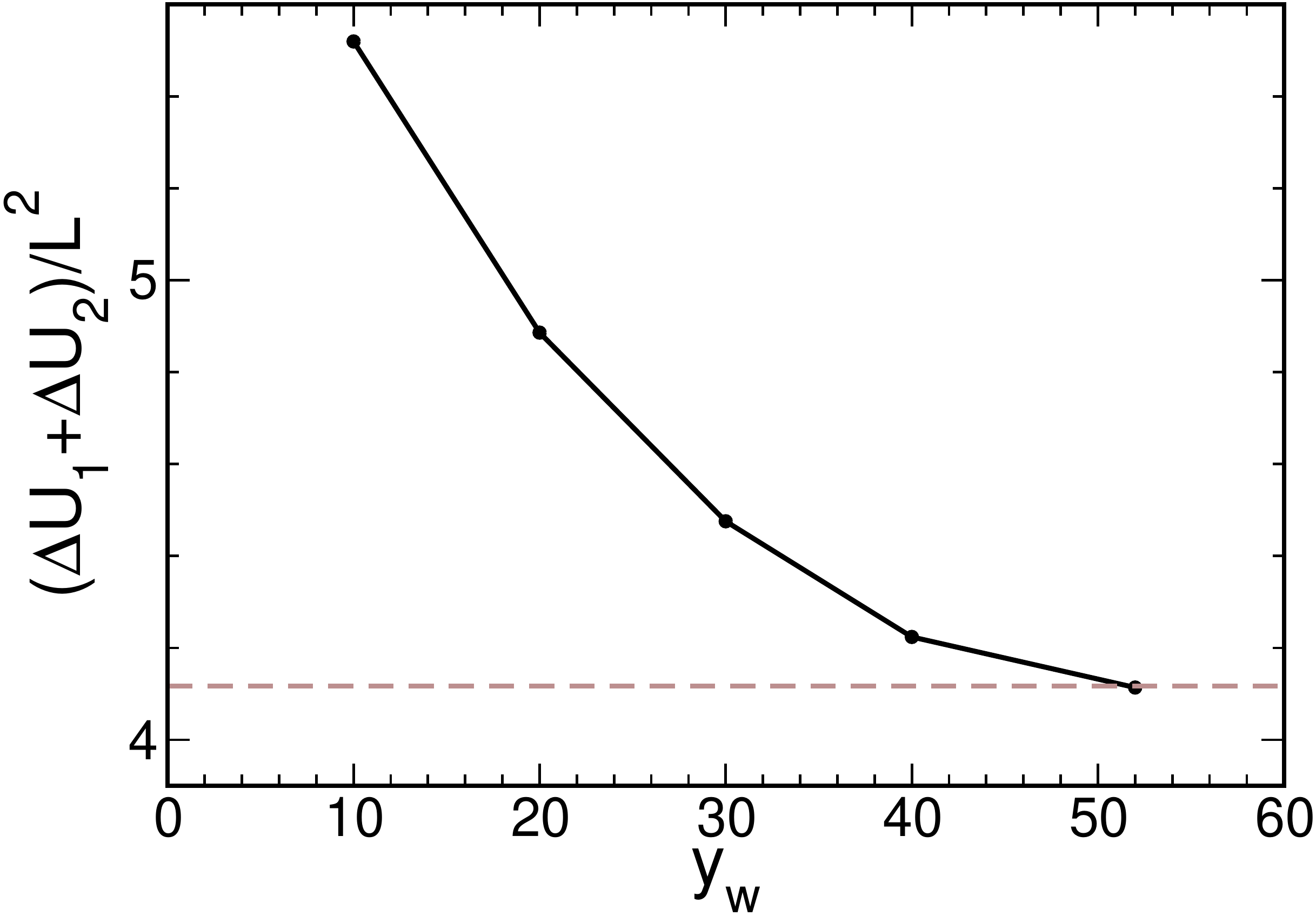}}
\caption{Row temperature profiles $\ovl{T}(y, \tau)$ and the stored energy under $A_{1}$ geometry, energy supply $\ovl{J_{U}}(A_{1}, E_{0}, \tau) = \ovl{J_{U}}(A_{1}, 20, 10000) = 0.002 \times L^2$. Temperature is measured in units of  $J/k_{\mathrm{B}}$. }
\label{A1}
\end{figure}

\begin{figure}
\centering
\subfigure[]{\includegraphics[scale=0.25]{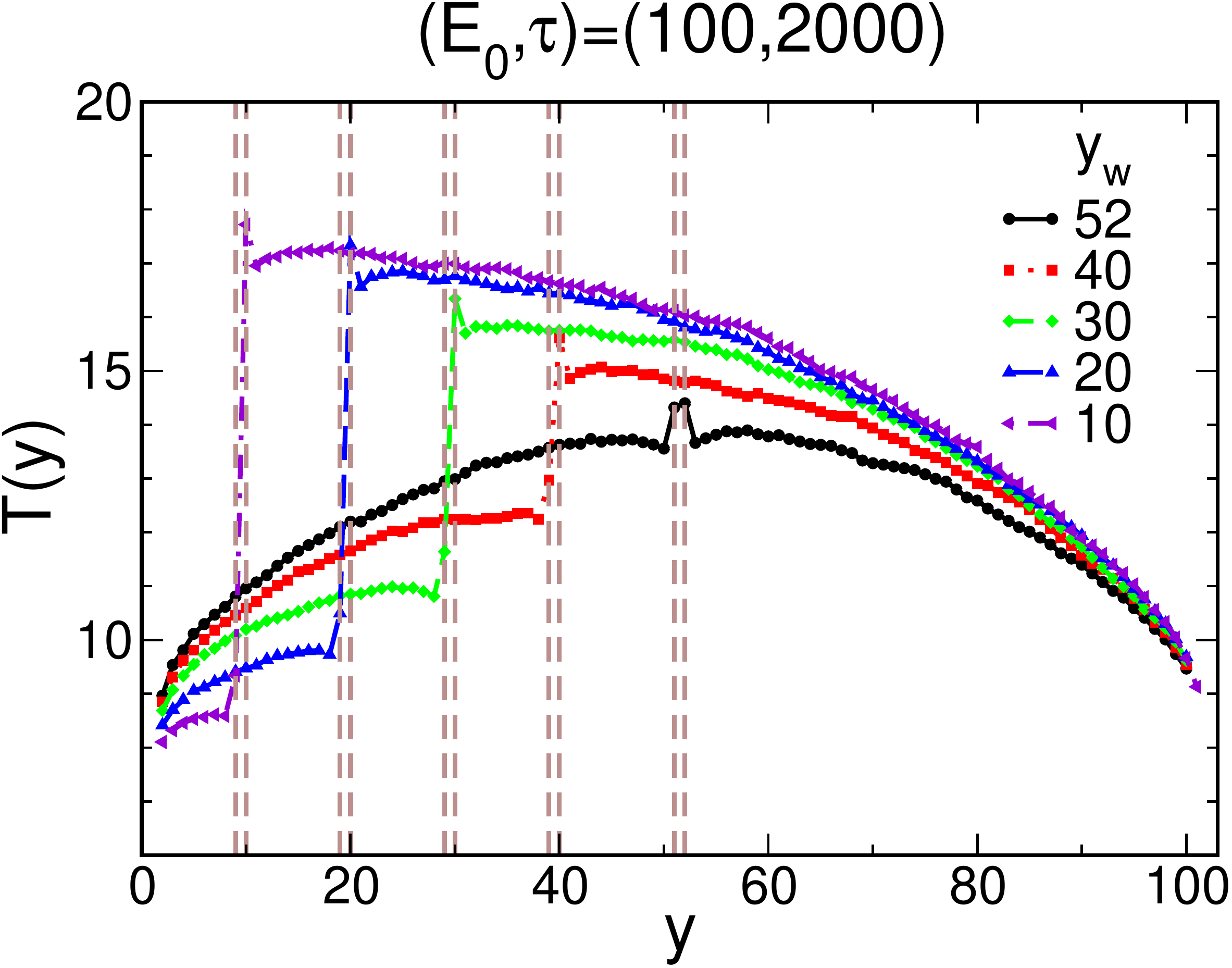}\label{A3_sub_Tx1}}
\subfigure[]{\includegraphics[scale=0.25]{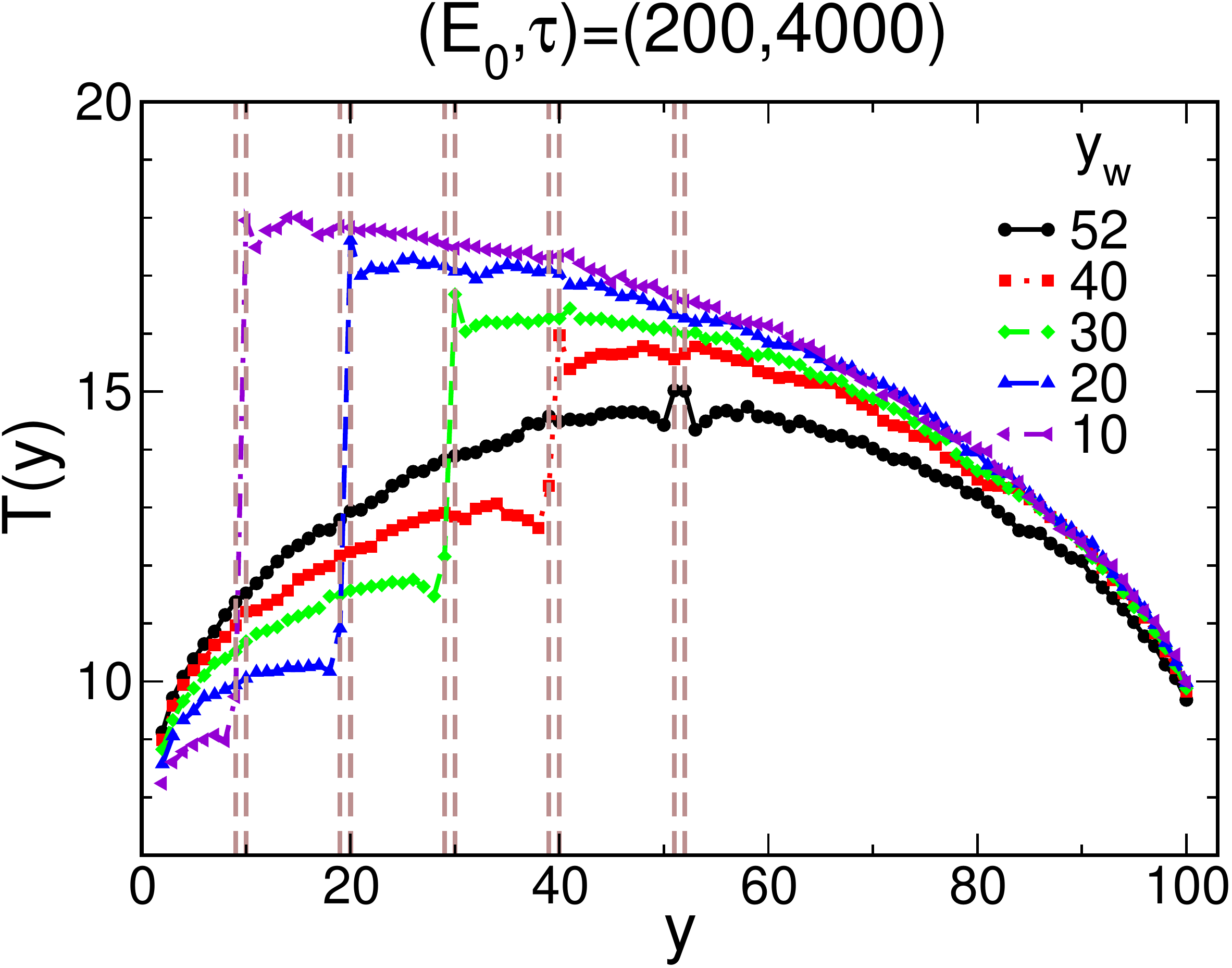}\label{A3_sub_Tx2}}
\subfigure[]{\includegraphics[scale=0.25]{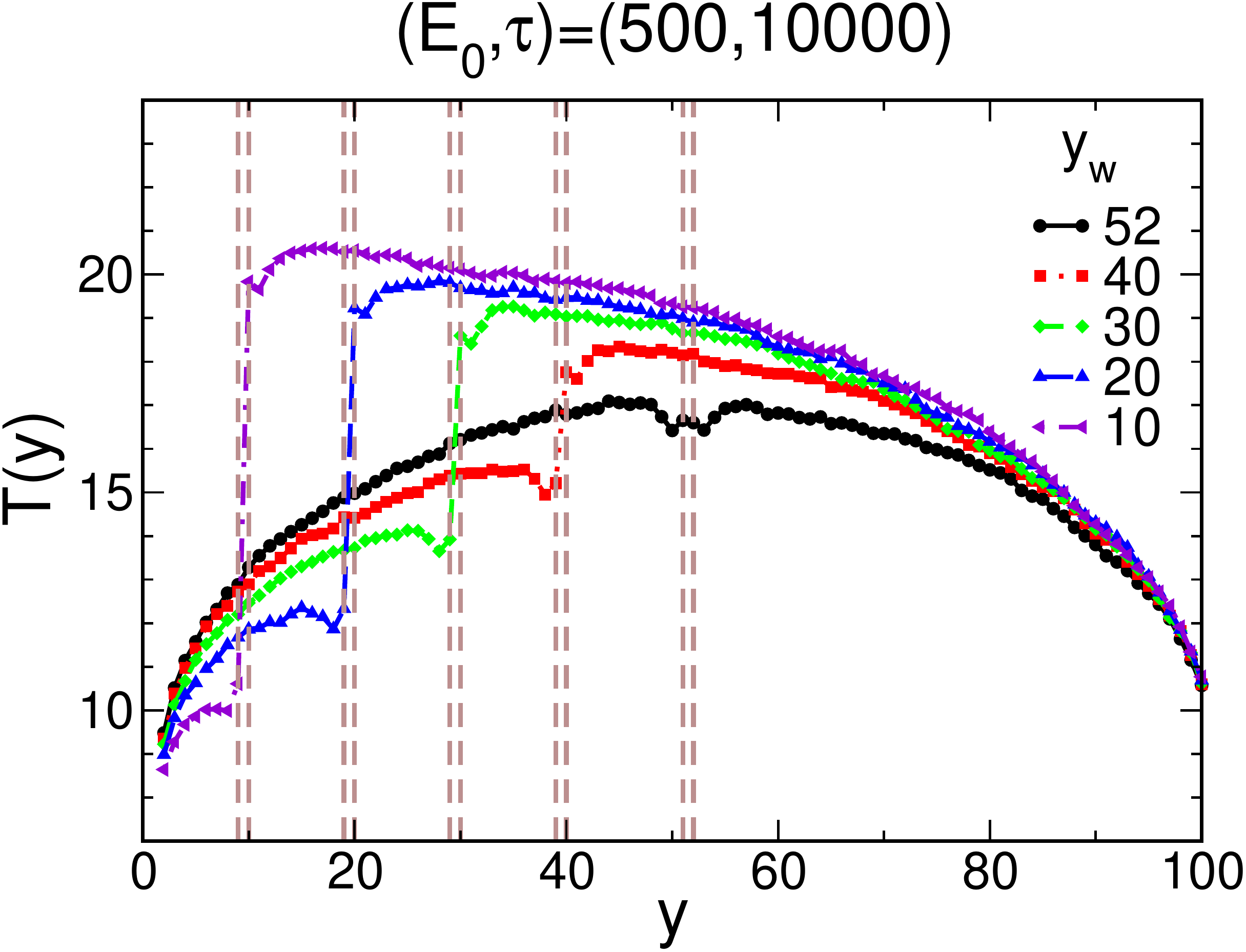}\label{A3_sub_Tx3}}
\subfigure[]{\includegraphics[scale=0.25]{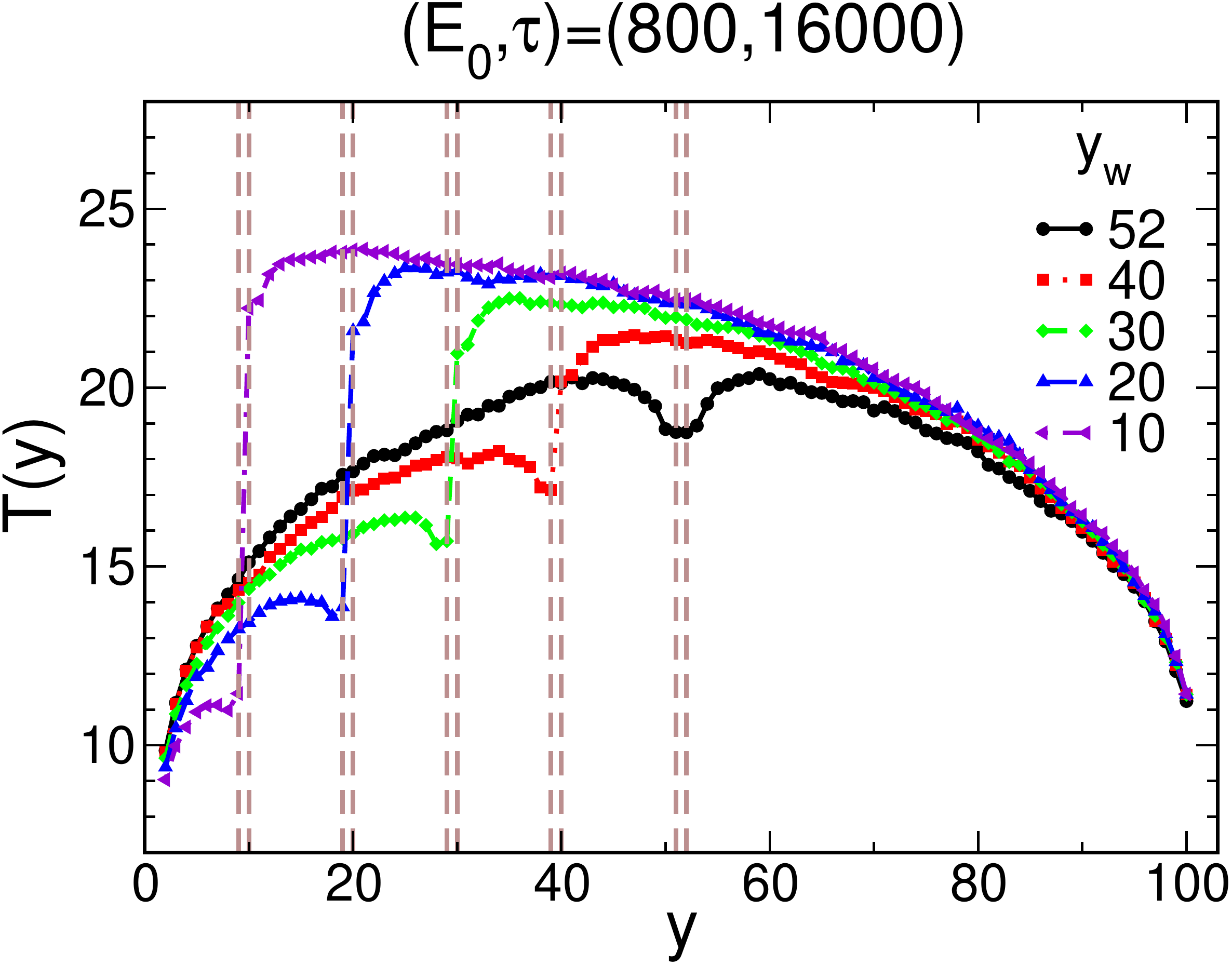}\label{A3_sub_Tx4}}
\subfigure[]{\includegraphics[scale=0.25]{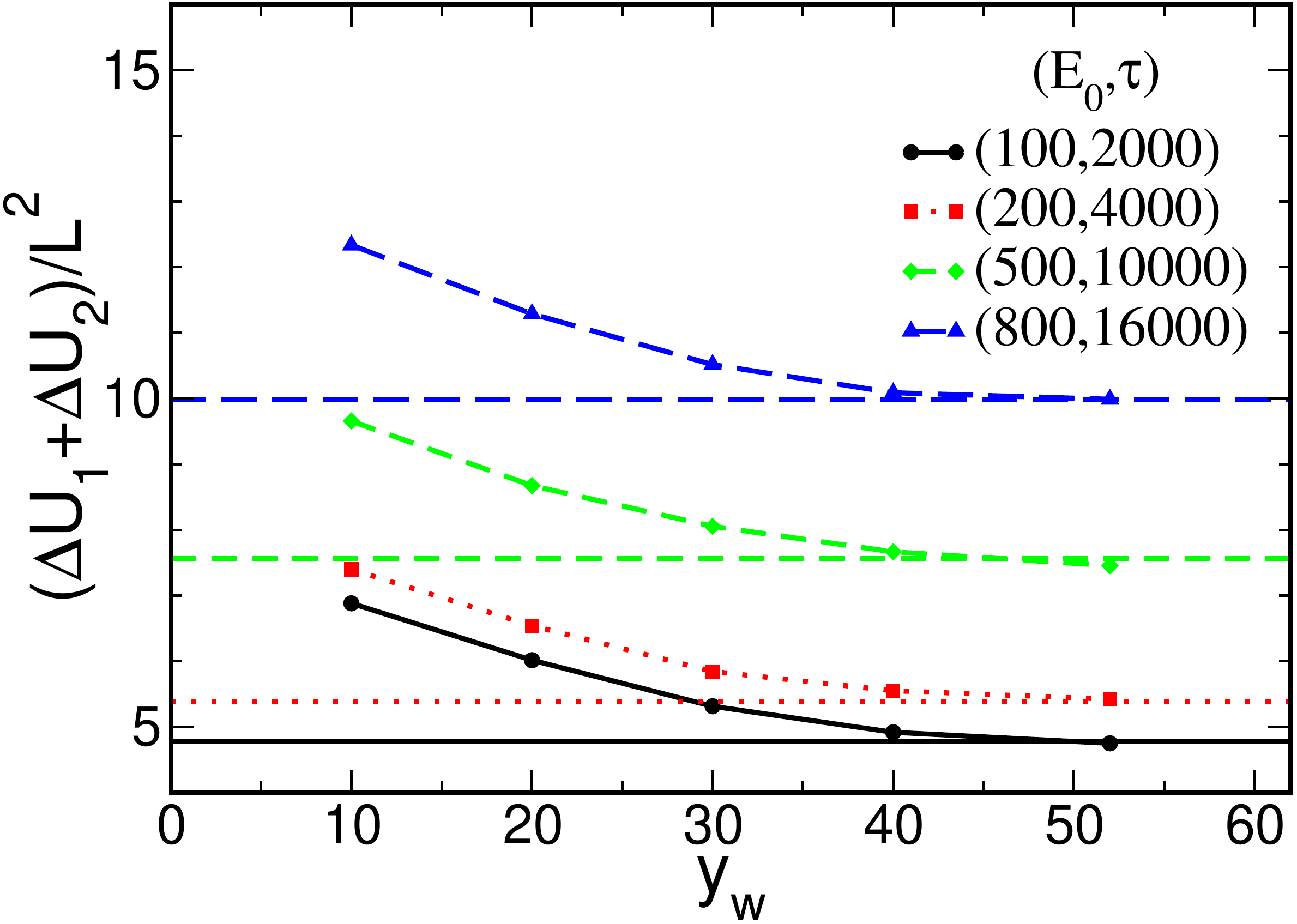}\label{A3_sub_dU}}
\caption{Row temperature profiles $\ovl{T}(y, \tau)$ and the stored energy under $A_{3}$ geometry. Averaged total  flux is fixed at $\ovl{J_{U}}(A_{3}, E_{0}, \tau) = 0.002 \times L^2$. Panels $(a)-(d)$ show results for  $\ovl{T}(y)$ against $y$ for four pairs of $(E_{0}, \tau)$. Corresponding energy amplitude and the period are indicated below each graph. Positions of the constrain are  denoted with dashed lines. Panel $(e)$ shows the stored energy for different pairs of $(E_{0}, \tau)$. Horizontal  lines correspond to the stored energy without constrain. Temperature is measured in units of  $J/k_{\mathrm{B}}$. }
\label{A3-Ty-same-protocol-diff-wall}
\end{figure}

For geometry $A_{1}$, we fixed the total flux with an energy supply of $\ovl{J_{U}}(A_{1}, E_{0}, \tau) = \ovl{J_{U}}(A_{1}, 20, 10000) = 0.002 \times L^2$. To observe the temperature change across the wall, we plotted the averaged 
row temperature $\ovl{T}(y, \tau) = \sum_{x}\ovl{T}(x,y, \tau)/L$ against $y$. Results of $\ovl{T}(y, \tau)$ and the stored energy with respect to different $y_{w}$ are shown in Fig.\ref{A1}. 
For geometry $A_{3}$, we fixed the total flux $\ovl{J_{U}} = 0.002 \times L^2$ and adjust four pairs of $(E_{0}, \tau)$. Fig.\ref{A3-Ty-same-protocol-diff-wall} shows the row temperature profiles for each pair of $(E_{0}, \tau)$ (Fig.\ref{A3-Ty-same-protocol-diff-wall}$(a)-(d)$) and the stored energy in all situations (Fig.\ref{A3_sub_dU}). Similar to the results for geometry $A_{1}$, under each protocol the stored energy of the periodic steady state in the unconstrained system is equal to that where the constrain separates the system into two equal parts. It is also lower than all other partitions.
This result is in accordance with the conjecture formulated in Ref.~\cite{holyst2019flux}.

\section{Conclusion}
In this paper, we have studied periodic steady states of a lattice system under different protocols of local cyclic energy supply using simulation. Combining the Metropolis dynamics and the deterministic dynamics  has allowed  us to specify the position for energy input, without assuming a temperature profile, and  to achieve heat transportation through temperature gradient.
We have  manipulated the geometry, amplitude and period of the protocol, while keeping the total flux constant and  compared the resulting  temperature profile and the stored energy of the periodic steady states.
Results show that the system reaches different steady states depending on the details of the protocol. 
This implies that the energy storage depends sensitively on  the mode of energy transfer into the system, which agrees with  findings  of  Ref.~\cite{holyst2019flux} for ideal gas and Lennard Jones fluids.
In Ref.~\cite{holyst2019flux} the modes of energy transfer were constant in time whereas in the present paper  we have employed cyclic energy supply and, therefore, we could examine the energy storage
in the periodic steady states.  Periodic external drives occur in many situations in physics and biology \cite{glass1988clocks, rikvold1998kinetic, tome1990dynamic}. 
We have observed that the system stores more energy through large and rare energy delivery, comparing to small and frequent delivery.  
Besides periodicity, also the  geometry of the energy supply  plays  an important role for the energy storage. 
In order  to store the same amount of energy  at the same  period  and  total energy flux, one has to supply much more energy  if the  delivery area small   than  if it is  large.
The temperature field in the periodic state depends also on the shape of the lattice. In this context, it would be interesting to explore different shapes of the system and also different lattice types.

We have also introduce a horizontal adiabatic wall as a constrain into the system. We have compared stored energy with the unconstrained case and found that the stored energy is minimized in this periodic steady states, which is also the state where the wall separate the system into two equal subsystems. This result further supports the hypothesis proposed by a recent study, where the quantity $\mathcal{T} \equiv \Delta U/J_{U}$, which links the stored energy $\Delta U$ with the energy flux, is minimized in steady states. 
Further studies exploring this quantity in  various examples of NESSs  would shed  light on  generality of  this hypothesis.  
 For example, the numerical approach that we use in our paper can be applied to systems with long-ranged interactions - see  Ref.~\cite{lederhendler2010long}.
 The energy storage in such system is an interesting issue, which we
shall consider in future studies.


\begin{acknowledgements}
The work of Y.Z. was partially supported by
the Polish National Science Centre (Harmonia Grant No.
2015/18/M/ST3/00403).
\end{acknowledgements}

\bibliographystyle{bibgen}
\bibliography{spin_rev}

\end{document}